\documentclass[PAPER, cernpreprint, atlasdraft=false, USenglish, texmf, orcidlogo]{atlasdoc}
\usepackage{atlaspackage}
\usepackage{atlasbiblatex}

\usepackage{atlasphysics}
\graphicspath{{logos/}{figures/}}

\usepackage{multirow}

\AtlasTitle{Measurement of the azimuthal anisotropy of charged particles in $\snn=5.36$~\TeV\ $^{16}$O$+^{16}$O and $^{20}$Ne$+^{20}$Ne collisions with the ATLAS detector}

\AtlasAbstract{%
This paper presents the first measurements of the azimuthal anisotropy
coefficients $v_{n}$, which quantify the $n^{\mathrm{th}}$-order
Fourier modulation of charged-particle azimuthal distributions,
for $n=2$--4 in $\sqrt{s_{\mathrm{NN}}}=5.36$~\TeV\
$\mathrm{^{16}O}+\mathrm{^{16}O}$ and $\mathrm{^{20}Ne}+\mathrm{^{20}Ne}$
collisions recorded with the ATLAS detector at the Large Hadron Collider in 2025.
The $v_{n}$ coefficients are measured as a function of transverse momentum (\pt),
collision centrality, and event multiplicity.
They are extracted using two complementary methods: two-particle correlations
with a template-fit subtraction of short-range non-flow contributions,
and four-particle subevent cumulants, which intrinsically suppress non-flow
effects and provide sensitivity to flow fluctuations.
The results show a clear hierarchy $v_{2} > v_{3} > v_{4}$ and a non-monotonic
dependence on \pt, reaching a maximum around $2$~\GeV, consistent with trends
observed in heavy-ion collisions.
Detailed comparisons between the two collision systems reveal an
enhanced $v_{2}$ in central $\mathrm{^{20}Ne}+\mathrm{^{20}Ne}$ collisions,
consistent with theory expectations based on the predicted prolate
deformation of neon nuclei, in contrast to the slightly tetrahedral structure
predicted for oxygen.
The four-particle cumulant results highlight strong event-by-event fluctuations
and provide the greatest sensitivity to nuclear shape effects.
These measurements can place new constraints on the initial geometry and the
hydrodynamic response in light-ion collisions, offering valuable input for
models of nuclear structure.
}

\AtlasRefCode{HION-2025-02}

\PreprintIdNumber{CERN-EP-2025-200}

\HEPDataRecord{ins2967110}

\AtlasJournalRef{Phys.\ Rev. \ C. 113 (2026) 045205}
\AtlasDOI{10.1103/xqxz-8bhf}

\hypersetup{pdftitle={ATLAS document},pdfauthor={The ATLAS Collaboration}}

\newcommand*{\OO}{O$+$O}
\newcommand*{\NeNe}{Ne$+$Ne}
\newcommand*{\pPb}{\textit{p}$+$Pb}
\newcommand*{\PbPb}{Pb$+$Pb\xspace}
\newcommand*{\XeXe}{Xe$+$Xe\xspace}

\newcommand{\papertype}{paper\xspace}
\newcommand{\fcalsum}{\ensuremath{\Sigma E_\mathrm{T}^\mathrm{FCal}}\xspace}
\providecommand*{\npart}{\ensuremath{N_{\mathrm{part}}}}

\providecommand*{\snn}{\ensuremath{\sqrt{s_{\text{NN}}}}}

\providecommand*{\deta}{\ensuremath{\Delta\eta}}
\providecommand*{\dphi}{\ensuremath{\Delta\phi}}
\providecommand*{\phia}{\ensuremath{\phi^{a}}}
\providecommand*{\phib}{\ensuremath{\phi^{b}}}
\providecommand*{\etaa}{\ensuremath{\eta^{a}}}
\providecommand*{\etab}{\ensuremath{\eta^{b}}}
\providecommand*{\pta}{\ensuremath{p_{\mathrm{T}}^{a}}}
\providecommand*{\ptb}{\ensuremath{p_{\mathrm{T}}^{b}}}
\providecommand*{\ptab}{\ensuremath{p_{\mathrm{T}}^{a,b}}}

\newcommand{\ctwo}{\ensuremath{C(\Delta\eta,\Delta\phi)}\xspace}
\newcommand{\ctwophi}{\ensuremath{C(\Delta\phi)}\xspace}
\newcommand{\tpc}{2PC}

\newcommand{\vn}{\ensuremath{v_n}}
\newcommand{\vnn}{\ensuremath{v_{n,n}}}

\newcommand{\vnfour}{\ensuremath{v_n\{4\}}}
\newcommand{\vntwo}{\ensuremath{v_n\{2\}}}
\newcommand{\vtwo}{\ensuremath{v_2}}
\newcommand{\vthree}{\ensuremath{v_3}}
\newcommand{\vfour}{\ensuremath{v_4}}

\newcommand{\nchrec}{\ensuremath{N_\mathrm{ch}^\mathrm{rec}}}

\newcommand{\vntpc}[1][n]{\ensuremath{v_{#1}^{\mathrm{2PC}}\{2\}}}
\newcommand{\vntf}[1][n]{\ensuremath{v_{#1}^{\mathrm{sub}}\{2\}}}

\providecommand*{\cperi}{\ensuremath{C^{\mathrm{periph}}(\Delta\phi)}}

\newcommand{\AtlasCoordFootnote}{%
ATLAS uses a right-handed coordinate system with its origin at the nominal interaction point (IP)
in the center of the detector and the $z$-axis along the beam pipe.
The $x$-axis points from the IP to the center of the LHC ring,
and the $y$-axis points upwards.
Cylindrical coordinates $(r,\phi)$ are used in the transverse plane,
$\phi$ being the azimuthal angle around the $z$-axis.
The pseudorapidity is defined in terms of the polar angle $\theta$ as $\eta = -\ln \tan(\theta/2)$.}

\usepackage{tikz}     %

\begin{document}

\maketitle

\section{Introduction}
\label{sec:intro}
A hot and dense state of nuclear matter in which the relevant degrees
of freedom are strongly coupled quarks and gluons is known as the
quark–gluon plasma (QGP).
This state can be created transiently in collisions of heavy nuclei at
high-energy colliders, such as the Relativistic Heavy Ion Collider
(RHIC) and the Large Hadron Collider (LHC)~\cite{
Adams:2005dq,Back:2004je,Adcox:2004mh,Arsene:2004fa,
HION-2011-05,HION-2016-06,HION-2019-03,
CMS-HIN-10-002,CMS-HIN-18-001,
ALICE:2010suc, ALICE:2018lao}.
The QGP produced in heavy-ion collisions undergoes collective expansion
driven by strong pressure
gradients, which convert initial-state spatial anisotropies into
momentum anisotropies of the final-state hadron
distribution~\cite{Adams:2005dq,Back:2004je,Adcox:2004mh,
Ollitrault:1992bk}.
This phenomenon, commonly referred to as collective flow, is well
described by nearly inviscid relativistic hydrodynamics and
constitutes a key signature of QGP
formation (see Ref.~\cite{Busza:2018rrf} and references therein).

The anisotropy of particle distributions in heavy-ion collisions is
quantitatively characterized by a Fourier series in the azimuthal
angle $\phi$~\cite{Alver:2010dn}:
\begin{equation}
\frac{dN}{d\phi} \propto 1 + 2 \sum_{n=1}^{\infty} v_n
\cos\left(n(\phi - \Psi_n)\right) ,
\label{eq:single}
\end{equation}
where $v_n$ and $\Psi_n$ represent the magnitude and orientation of the
$n^\text{th}$-order anisotropy, respectively.
The $v_n$ are commonly referred to as ``flow harmonics,'' while the
$\Psi_n$ are referred to as ``event-plane angles.''
The $v_n$ depend on transverse momentum $\pt$,
pseudorapidity\footnote{\AtlasCoordFootnote} ($\eta$), and event
multiplicity, and fluctuate event-by-event
(EbE)~\cite{HION-2011-01,HION-2012-10,CMS-HIN-16-019,ALICE:2018rtz}.
These EbE fluctuations arise primarily from variations in the initial
geometry and energy density of the nuclear overlap region, which are
driven by the fluctuating positions of nucleons and by subnucleonic
structure, collectively referred to as geometric fluctuations.
Among these coefficients, elliptic flow ($v_2$) is the largest due to
the lenticular geometry of the average overlap region.
This is typically followed by triangular flow ($v_3$), which typically
has no contribution from the average geometry
and is therefore generated entirely by EbE geometric fluctuations.
For higher orders ($n \geq 4$), the flow harmonics are influenced not
only by the initial geometry but also by non-linear mode coupling
from lower-order harmonics, such as $v_{4}$ from $v_{2}^{2}$ and
$v_{5}$ from $v_{2}v_{3}$~\cite{HION-2012-03}.
These contributions make the higher-order harmonics sensitive to both
the geometry and the medium’s collective response.
Extensive studies of $v_n$ and their EbE fluctuations have provided
important constraints on the initial-state geometry and on transport
properties of the QGP in nuclear collisions, such as the shear
viscosity to entropy density ratio $\eta/s$
(see Ref.~\cite{Busza:2018rrf} and references therein).

The observation of collective flow was initially regarded as an
exclusive signature of QGP formation in collisions of heavy nuclei,
such as Au+Au or Pb+Pb systems.
However, experiments at the LHC and the RHIC have revealed large $v_n$
signals in much smaller collision systems, including $pp$ and $p$+Pb
at the LHC~\cite{ALICE:2012eyl,HION-2012-13,HION-2015-09,CMS-HIN-16-010}
and $p$/$d$/$^3$He+Au at
the RHIC~\cite{STAR:2023wmd,STAR:2022pfn,PHENIX:2021ubk,PHENIX:2018lia}.
Remarkably, the observed $v_n$ hierarchy in these small systems follows
patterns consistent with expectations based on geometrical
differences between their initial conditions.
Detailed theory studies suggest that the measured \vn\ in small
systems can be explained by the collective expansion of the matter
produced in the collision, driven by the shape and fluctuations of
the initial overlap region~\cite{Weller:2017tsr,Bzdak:2013zma}.
Nevertheless, substantial uncertainties remain regarding the precise
nature and properties of the produced matter in small collision systems.
These uncertainties arise primarily from the incomplete understanding
of the initial conditions in small systems, which are sensitive not
only to the spatial distribution of nucleons within nuclei but also
to the internal structure of individual
nucleons~\cite{STAR:2023wmd,Mantysaari:2016ykx}.
Consequently, precise characterization of the initial conditions in
small systems is essential to improve the extraction of medium
properties such as shear viscosity, pre-hydrodynamic evolution
effects, and initial-state momentum
anisotropies~\cite{Giacalone:2017dud,Ambrus:2024eqa,Noronha:2024dtq}.

To address these uncertainties, a promising experimental strategy is
to compare $v_n$ harmonics in collisions of nuclei with similar mass
numbers but distinct nucleon
arrangements~\cite{Citron:2018lsq,Jia:2022ozr,STAR:2021mii}.
Such systems are expected to produce matter with comparable bulk
properties and similar final-state collective responses.
Therefore, the ratios of $v_n$ between the two systems are expected to
be primarily sensitive to differences between their initial
conditions~\cite{Giacalone:2021uhj}.
Analogous comparisons have previously been used to study the influence
of nuclear structure in large collision
systems~\cite{HION-2021-01,ALICE:2024nqd,STAR:2024wgy}.
A key advantage of applying this approach in small collision systems is
that the initial-state differences are governed by nucleon
distributions that can, in principle, be calculated using
state-of-the-art \textit{ab initio} nuclear structure
theories~\cite{Hergert:2020bxy}, providing a theoretical foundation
for interpreting experimental observations.

An ideal pair of collision systems for such comparative studies is
$^{16}$O+$^{16}$O and $^{20}$Ne+$^{20}$Ne.
Low-energy experiments and theory calculations indicate that
$^{16}$O has a near-spherical tetrahedral shape, whereas $^{20}$Ne
is predicted to consist of a $^4$He cluster ``orbiting'' an $^{16}$O
core~\cite{cluster,PTP.42.555}.
Hydrodynamic simulations incorporating nuclear configurations from
\textit{ab initio} calculations predict significant enhancements in
$v_2$ and $v_3$ for central $^{20}$Ne+$^{20}$Ne collisions compared
with central $^{16}$O+$^{16}$O
collisions\footnote{For simplicity, $^{16}$O and $^{20}$Ne are hereafter denoted as O and
Ne, respectively.}~\cite{Giacalone:2024luz}.
Furthermore, nuclear structure effects are also predicted to modify the
EbE fluctuations of $v_2$ and $v_3$ between the two
systems~\cite{Wang:2024ulq}, providing additional discriminatory
power for theory models.

This paper presents measurements of anisotropic flow coefficients
$v_n$ for $n=2$ to 4 in $\snn = 5.36$~\TeV\ \OO\ and \NeNe\ collisions
recorded by the ATLAS detector at the LHC in 2025.
The analysis employs both two- and four-particle correlation
methods~\cite{HION-2015-09,HION-2017-02}, which probe different
moments of the $v_n$ distributions.
These measurements provide valuable constraints on the final-state
collective response and on the properties of the produced medium in
each system individually, while their comparison offers unique
insights into the initial-state geometry and nucleon distributions
within these light nuclei.
After this manuscript was submitted for publicaion, related measurements
from the ALICE and CMS Collaborations~\cite{AliceLightIonFlow,CMS-HIN-25-009}
were also submitted.

\section{Experimental configuration}
\label{sec:detector}

The ATLAS detector~\cite{PERF-2007-01,GENR-2019-02} at the LHC covers
nearly the entire solid angle around the collision point.
It consists of an inner tracking detector surrounded by a thin
superconducting solenoid, electromagnetic and hadronic calorimeters,
and a muon spectrometer incorporating three large superconducting
air-core toroidal magnets.
The primary subsystems relevant to this study include the
inner detector (ID), the calorimeter and the trigger
and data acquisition infrastructure.

The ID is immersed in a \qty{2}{\tesla} axial magnetic field
and provides charged-particle tracking in the range \(|\eta| < 2.5\).
The high-granularity silicon pixel detector covers the vertex region
and typically provides four measurements per track,
the first hit generally being in the insertable B-layer (IBL).
It is followed by the SemiConductor Tracker (SCT), which usually
provides eight measurements per track.
These silicon detectors are complemented by the transition
radiation tracker (TRT), which enables radially extended
track reconstruction up to \(|\eta| = 2.0\).

The calorimetry system %
includes several components:
a liquid argon (LAr) electromagnetic calorimeter covering $|\eta| < 3.2$;
a steel–scintillator sampling hadronic calorimeter covering $|\eta| < 1.7$;
an additional LAr-based hadronic calorimeter for the region
$1.5 < |\eta| < 3.2$;
and forward LAr calorimeters (FCal) designed to measure both
electromagnetic and hadronic activity in the range $3.2 < |\eta| < 4.9$.
Forward neutrons produced from the breakup of nuclei in both hadronic and
electromagnetic interactions are measured by compact tungsten sampling
Zero Degree Calorimeters (ZDCs) positioned at $z=\pm140$ meters from
the ATLAS interaction point.

Event collection is handled by a two-tiered trigger
system~\cite{TRIG-2022-01}.
The first level (L1) is implemented through a combination of
custom hardware and programmable logic, while the high-level
trigger (HLT) uses  software algorithms to refine the selection
using more detailed detector information.

A software suite~\cite{SOFT-2022-02} is used in data simulation,
in the reconstruction and analysis of real and simulated data,
in detector operations, and in the trigger and data acquisition
systems of the experiment.

\section{Datasets, event and track selection}
\label{sec:datasets}

The \OO\ and \NeNe\ data used in this \papertype\ were collected in July 2025.
The datasets correspond to integrated
luminosities of 2~\inb\ for \OO\ and 0.5~\inb\ for \NeNe\ collisions,
respectively.
Minimum-bias events were selected with an L1 trigger based on the
TRT ``FastOR'' algorithm~\cite{ATL-INDET-PUB-2009-002},
that required at least one TRT azimuthal sector above threshold,
together with an HLT requirement of at least one reconstructed
charged-particle track.
To increase statistics for the highest-activity events, an additional
trigger was used that required the TRT FastOR at L1 and at least
290 reconstructed tracks with $\pt > 200$~\MeV\ at the HLT.
A similar trigger with a requirement of 100 reconstructed tracks with
$\pt > 200$~\MeV\ at the HLT was employed to supplement the
intermediate-activity region.
In addition, a trigger requiring an L1 calorimeter transverse energy
threshold of 20~\GeV, followed by at least one reconstructed track
with $\pt > 200$~\MeV\ at the HLT, was also included.

In the offline analysis, the $z$-position of the primary
vertex~\cite{ATL-PHYS-PUB-2015-026} was required to lie within
10~cm of the nominal interaction point (i.e. within 10 cm of the center of ATLAS).
In the recorded \OO\ and \NeNe\ data, there are sizable contributions
from pileup events, in which two or more inelastic collisions
occur in the same bunch crossing.
The majority of the pileup events are removed by requiring only a single
high-quality reconstructed vertex per event.
A high-quality vertex is defined as one that is well-constrained in position, having
a $z$-position variance of less than $0.02~\mathrm{mm}^2$;
such vertices have many associated tracks pointing to a common origin.
Additional background is suppressed by removing events with significantly
smaller track multiplicities than expected relative to the total transverse
energy recorded in the FCal (\fcalsum)~\cite{HION-2012-04}.
Similarly, correlations between the energy deposited in the ZDCs and \fcalsum\
are used to suppress additional pileup by rejecting events for which
the energy deposited in the ZDCs is significantly higher
than in the majority of collisions.
The estimated residual pileup after these selections is found to be
at most 0.2\%.

\begin{figure}
\centering
\includegraphics[width=0.99\linewidth]{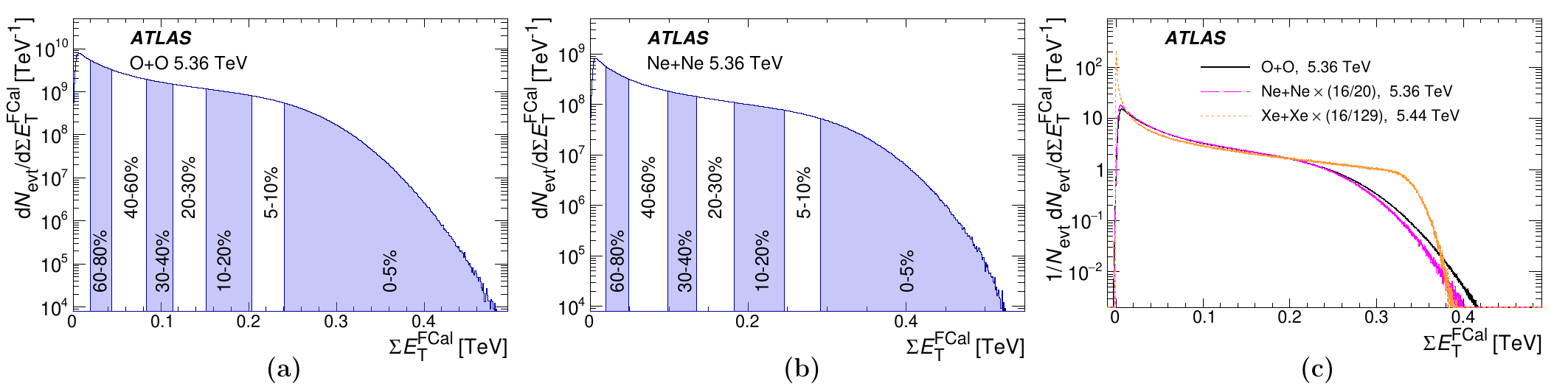}
\caption{
The \fcalsum\ distribution in
minimum-bias events, together with the thresholds
for a few centrality intervals,
for (a) \OO\ collisions and (b) \NeNe\ collisions.
(c) Comparison to Xe$+$Xe collisions~\cite{HION-2019-03},
where the \fcalsum\ of the \XeXe\ and \NeNe\ systems are
scaled by the relative number of nucleons to oxygen.
The \NeNe\ and \XeXe\ distributions in panel (c)
are normalized to have the same integral as the
\OO\ distribution above 30~\GeV.
}
\label{fig:centrality}
\end{figure}

As in earlier ATLAS studies of heavy-ion collisions, events are categorized
into centrality percentiles based on the
\fcalsum~\cite{HION-2011-01,HION-2011-05,HION-2019-03}.
To relate the \fcalsum\ distribution to the sampled fraction of the
total inelastic \OO\ and \NeNe\ cross-sections,
a Glauber-model--based calculation~\cite{Miller:2007ri,Loizides:2025ule}
is used to fit the data, to extract the fraction of events selected
above a minimum \fcalsum\ threshold, and to estimate the systematic
uncertainties on that fraction~\cite{HION-2011-01,HION-2011-05,HION-2019-03}.
Additionally, the Glauber model is used to extract primary collision
characteristics, such as the average number of participating nucleons,
$\langle\npart\rangle$, for each centrality interval.
The distribution of \fcalsum\ observed in data, along with the
threshold values defining various centrality intervals,
is illustrated in Figure~\ref{fig:centrality}.
Figure~\ref{fig:centrality}c shows the comparison of the \fcalsum\
distributions in \OO\ and \NeNe, to that in Xe$+$Xe collisions
(from Ref.~\cite{HION-2019-03}).
The comparison is done after scaling the \NeNe\ and \XeXe\
\fcalsum\ distributions by the number of nucleons relative to oxygen.
While the \XeXe\ distribution exhibits a sharper fall-off at a scaled
\fcalsum\ of $\sim0.34~\TeV$ and a narrower tail beyond this point,
the \OO\ and \NeNe\ distributions show a significantly more pronounced
high-\fcalsum\ tail.
This behavior reflects the larger relative event-by-event fluctuations
in particle production in the smaller \OO\ and \NeNe\ systems compared
to \XeXe.

\begin{figure}
\centering
\includegraphics[width=0.99\linewidth]{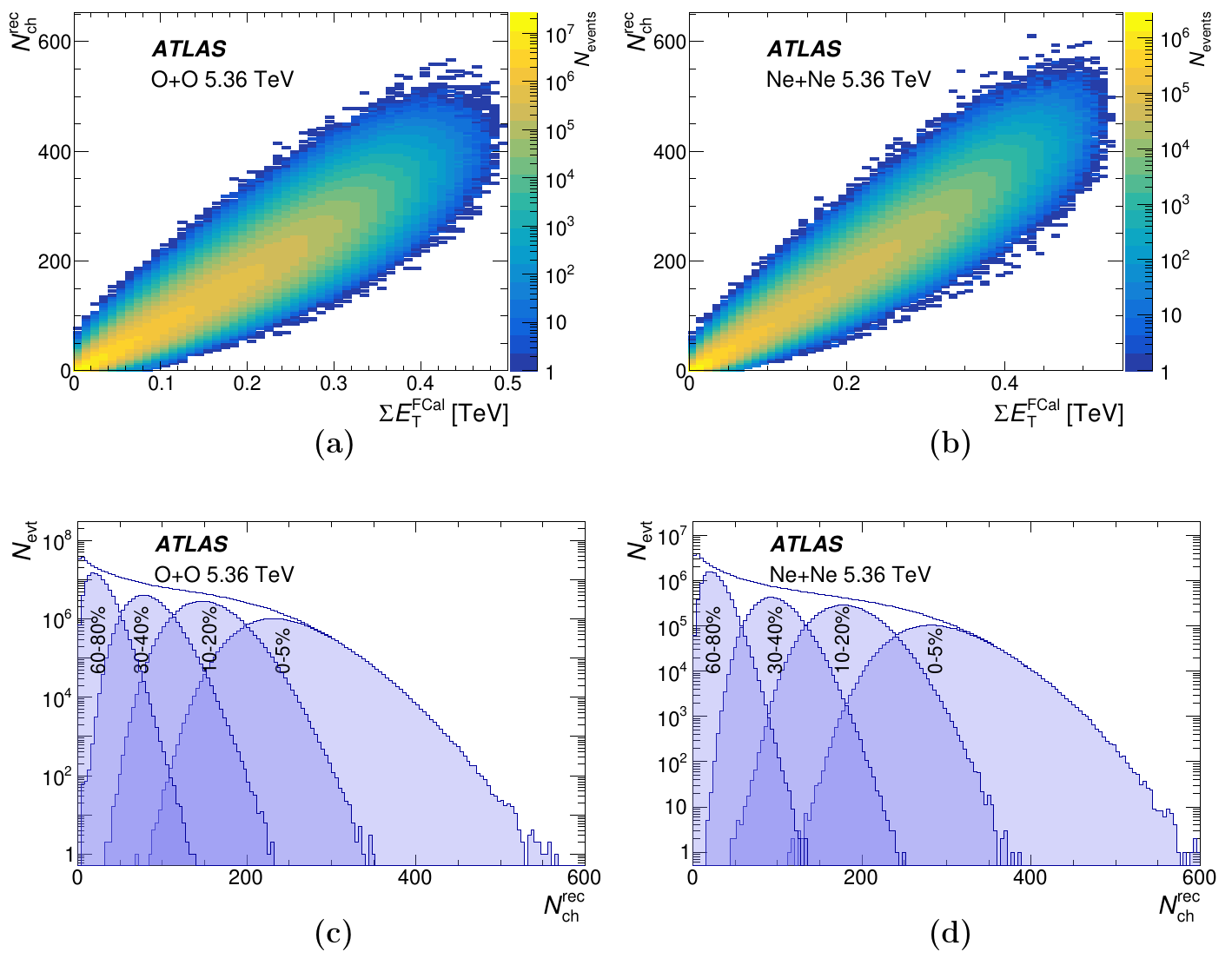}
\caption{
Correlation between \nchrec\ and \fcalsum\ in minimum-bias
events for (a) \OO\ and (b) \NeNe\ collisions, and
distributions of \nchrec\ in minimum-bias events for (c) \OO\
and (d) \NeNe\ collisions, including the distributions
for several centrality intervals.
}
\label{fig:nchrec}
\end{figure}

Charged-particle tracks and collision vertices are reconstructed from
hits in the ID using standard methods~\cite{PERF-2015-08}.
For the nominal analysis, the reconstructed tracks are required to
have $\pt > 0.5~\GeV$, $|\eta| < 2.5$ and at least one pixel hit,
with the additional requirement of a hit in the IBL when one
is expected\footnote{
A hit is expected if the extrapolated track crosses an active
region of a silicon-sensor module (pixel or SCT) that has not
been disabled, and a hit is said to be ``missing'' when it is
expected but not found.
If a track crosses a disabled module, then for the
purposes of hit counting, the disabled module is counted as a hit.
}.
If a hit in the IBL is not expected, then a hit is required in
the next-to-innermost pixel layer, if such a hit is expected.
The tracks are required to have at least six SCT hits.
To suppress secondary contributions, the transverse impact parameter of the track with
respect to the beam-line, $d_0$, and the longitudinal impact parameter
of the track relative to the primary vertex, $z_0\sin(\theta)$,
are required to satisfy
$|d_0| < 1.5$~mm and $|z_0\sin(\theta)| < 1.5$~mm.
The quantity \nchrec\ is defined as the number of charged-particle tracks
in an event that satisfy these selection criteria and have $\pt > 0.5$~\GeV.
An alternate set of more restrictive selections are used to evaluate
systematic uncertainties in the measurement.
For these ``tight'' selections, the number of pixel and SCT hits
are raised to two and eight respectively,
a requirement of at most one missing hit in the SCT is imposed,
and the $d_0$ and $z_0\sin(\theta)$ impact parameter selections are decreased to 1~mm.
Furthermore, the $\chi^2$ per degree of freedom of the reconstructed
track trajectory is required to be less than 6.

Figures~\ref{fig:nchrec}a and \ref{fig:nchrec}b show the correlation
between \fcalsum\ and \nchrec\ in minimum-bias \OO\ and \NeNe\
events, respectively.
These two measures of event activity are found to be well
correlated.
Figures~\ref{fig:nchrec}c and \ref{fig:nchrec}d show the \nchrec\
distributions in minimum-bias \OO\ and \NeNe\ events, together with
the distributions for several centrality intervals from
Fig.~\ref{fig:centrality}.

To study the detector performance,
a sample of 7 million minimum-bias \OO\ Monte Carlo (MC) events was generated
using the HIJING event generator version 1.38b~\cite{Wang:1991}.
Since HIJING does not have any intrinsic mechanism to generate flow,
the latter is added after the initial particle generation step using an
``afterburner'' procedure~\cite{Jia:2013tja}, which slightly shifts the $\phi$
positions of generated particles to mimic flow.
The generated sample was then passed through a full simulation of the ATLAS
detector~\cite{SOFT-2010-01} using \GEANT~\cite{Agostinelli:2002hh}, and the MC events
were reconstructed by the same algorithms as the data.
The reconstructed particles in the MC events were used to calculate the
reconstruction efficiency -- the fraction of the generated charged
particles that are successfully reconstructed and selected
-- as a function of \pt\ and $\eta$, and denoted by $\epsilon(\pt,\eta)$ below.
With the criteria imposed in this analysis,
the efficiency at $\pt=1~\GeV$ varies between $\sim 65\%$ at $|\eta|=2.5$
to $\sim75\%$ at $|\eta|=2$ and $\sim85\% $ at $|\eta|=0$.
At mid-rapidity ($|\eta|<1$) the efficiency is $\sim80\%$ at
$\pt=0.5~\GeV$ and increases to $\sim90\%$ at a \pt\ of 5~\GeV.
The rate of fake tracks, tracks that do not correspond to
any generated particle, denoted by $f(\pt,\eta)$,
is also estimated from the MC and stays below
$\sim2\%$ across the \pt\ and $\eta$ ranges
used in this measurement.
The reconstructed event multiplicity \nchrec\ is corrected for
reconstruction efficiency and fake tracks by weighting each track
with:
\begin{eqnarray}
\frac{1-f(\pt,\eta)}{\epsilon(\pt,\eta)},
\label{eq:effcorr}
\end{eqnarray}
where the numerator accounts for fake-track removal and the denominator
for the reconstruction efficiency.
This efficiency- and fake-corrected multiplicity is referred to as
$N_\mathrm{ch}$.

\section{Methodology}
\label{sec:analysis}

Due to the limited multiplicity per event, the \vn\ cannot be
reliably measured on an EbE basis.
Instead, the flow harmonics \vn\ are estimated from multi-particle
correlations, which average over many events and provide access to
different moments of the EbE \vn\ distributions.
However, correlations unrelated to collective flow can contaminate the
measurement and must be suppressed or removed.
For example, jet production and resonance decays produce strongly
correlated collinear particles ($|\deta|, |\dphi|<1$), and
weaker but equally important back-to-back correlations at $\dphi
\approx \pi$ that persist even at large $\deta$.
These few-particle, non-global correlations are referred to as
``non-flow.''
This section describes two approaches for estimating flow harmonics: the two-particle
correlations (\tpc) method, including its improved template-fit implementation,
and the multi-particle cumulant method.
In the \tpc\ approach, a template fit is used to subtract non-flow contributions
based on the lowest-multiplicity events, while the multi-particle cumulant
method applies a subevent technique to suppress non-flow effects.

\subsection{Two-particle correlations and template fit}
\label{sec:2pc}

\begin{figure}
\centering
\includegraphics[width=0.99\textwidth]{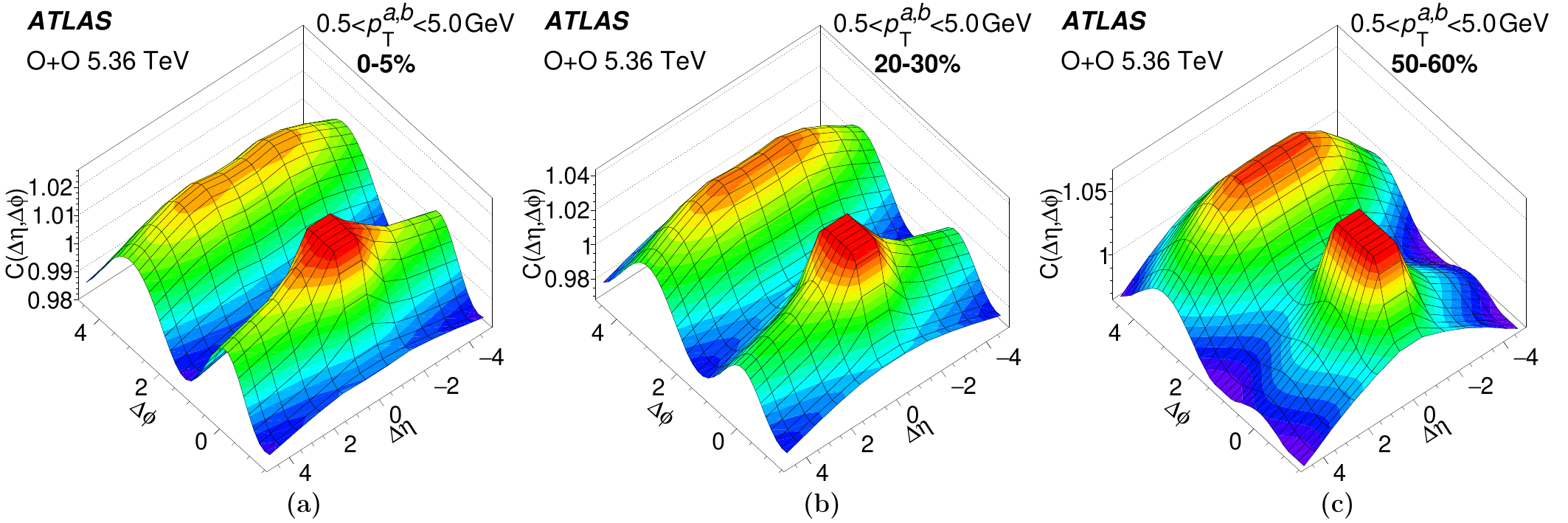}
\caption{
Two-particle $\deta-\dphi$ correlations in \OO\ collisions for
the (a) 0--5\%, (b) 20--30\%, and (c) 50--60\% centrality intervals.
The plots are for $0.5<\ptab<5$~\GeV.
The distributions are truncated along the $z$-axis,
to suppress the peak at $\deta=\dphi=0$,
and are plotted over $|\deta|<4.5$ to avoid statistical
fluctuations at larger $|\deta|$.
}
\label{fig:2d_2pcs2}
\end{figure}

The \tpc\ method has been widely used for flow measurements at the RHIC
and the LHC~\cite{Adare:2008ae,HION-2011-01,HION-2012-10,HION-2012-13,
HION-2013-04,HION-2014-03,HION-2015-09,HION-2016-01,HION-2016-06,
CMS-HIN-11-006,CMS-HIN-11-001,CMS-HIN-12-011,CMS-HIN-18-001,
ALICE:2011svq,ALICE:2023gyf}.
Correlations between pairs of charged particles are studied as a function
of their relative pseudorapidity, $\deta = \etaa - \etab$,
and relative azimuthal angle, $\dphi = \phia - \phib$.
The indices $a$ and $b$ denote the two particles in the pair,
whose kinematic selections may differ.
To account for detector acceptance effects, the correlation
function is defined as the ratio of the ``same-event'' pair distribution $S$,
where both particles are taken from the same event, to the
``mixed-event'' distribution $B$, where they are taken from
different events~\cite{HION-2011-01}:
\begin{eqnarray*}
\ctwo = \frac{S(\deta,\dphi)}{B(\deta,\dphi)}.
\end{eqnarray*}
The same-event distribution contains both genuine physical correlations
and contributions from detector acceptance,
inefficiencies and nonuniformities that are not related to the underlying physics.
The mixed-event distribution reflects only these non-physical effects,
so that their ratio isolates the genuine physical
correlations~\cite{Adare:2008ae}.
To ensure this, events used for the $B$ distribution are required
to have similar centrality (or multiplicity) and vertex position.
When constructing $S$ and $B$, corrections for track reconstruction
inefficiency and fake tracks are applied using a per-pair weight
(see Eq.~\ref{eq:effcorr}):
\begin{eqnarray*}
\frac{(1-f(\pta,\etaa))(1-f(\ptb,\etab))}{\epsilon(\pta,\etaa)\,\epsilon(\ptb,\etab)}.
\end{eqnarray*}
Examples of $\ctwo$ are shown in Figure~\ref{fig:2d_2pcs2}, normalized
such that the integral of $B(\deta,\dphi)$ matches that of
$S(\deta,\dphi)$ for $|\deta|>2$.
In all cases a prominent peak is observed at $\deta=\dphi=0$, arising
from short-range correlations such as jet fragmentation, resonance
decays, or Hanbury Brown–Twiss (HBT)
correlations~\cite{Lisa:2005dd}.
At large $\deta$, long-range correlations are visible both on the
near-side ($\dphi\sim0$) and the away-side ($\dphi\sim\pi$).
The near-side correlation, commonly referred to as the ``ridge,''
originates primarily from collective flow.
The away-side correlation receives contributions from both collective
flow and back-to-back dijets.

One-dimensional correlation functions, \ctwophi, are obtained by integrating
the $S$ and $B$ distributions over the range $2 < |\deta| < 5$:
\begin{eqnarray*}
\ctwophi = \frac{\int_{2}^{5} S(|\deta|,\dphi)\, d|\deta|}{\int_{2}^{5} B(|\deta|,\dphi)\, d|\deta|}
\equiv \frac{S(\dphi)}{B(\dphi)},
\end{eqnarray*}
where the $|\deta|>2$ requirement is imposed to suppress
non-flow correlations arising from the peak at $\deta=\dphi=0$
seen in Figure~\ref{fig:2d_2pcs2}~\cite{HION-2012-13,HION-2013-04,HION-2015-09}.
The \ctwophi\ are normalized to have an average value of unity.
Similar to the single-particle distribution (Eq.~\ref{eq:single}),
the \ctwophi\ is parametrized with a Fourier
series~\cite{HION-2011-01}:
\begin{eqnarray}
\ctwophi=C_{0}\left(1+2\Sigma_{n=1}^{\infty}\vnn(\pta,\ptb)\cos(n\dphi)\right).
\label{eq:2pc_fourier}
\end{eqnarray}

To suppress residual non-flow contributions that persist over $|\deta| > 2$,
primarily on the away side ($\Delta\phi \sim \pi$), a commonly used template-fit
procedure is employed~\cite{HION-2015-09,HION-2016-01,HION-2017-11,STAR:2023wmd}.
In this method, the shape of the non-flow component is estimated from
low-multiplicity (peripheral) events and assumed to be unchanged in
higher-multiplicity (more central) events.
For this analysis, the peripheral reference correlation, \cperi,
is constructed  from events of the same collision species (\OO\ or \NeNe)
with centrality greater than 80\%.
The measured correlation \ctwophi\ is then parameterized as the sum of this
non-flow reference and an azimuthally modulated pedestal, $C^\mathrm{ridge}(\Delta\phi)$,
that encodes the collective anisotropy:
\begin{align}
\ctwophi &= F\cperi +C ^\mathrm{ridge}(\Delta\phi),
\label{eq:tempfit_corr}
\end{align}
with
\begin{eqnarray}
C^\mathrm{ridge}(\Delta\phi) \equiv G\left[1+2\sum_{n=2}^{5}\vnn(\pta,\ptb)\cos(n\Delta\phi)\right].
\label{eq:tempfit_expansion}
\end{eqnarray}
The parameters $F$, $G$, and the \vnn\ are determined by the template fit, with $F$ and $G$
constrained such that the integrals of both sides of Eq.~\ref{eq:tempfit_corr}
are equal.
Fourier terms up to fifth order ($v_{2,2}$-$v_{5,5}$) are included
in the fit.

\begin{figure}
\centering
\includegraphics[width=0.99\linewidth]{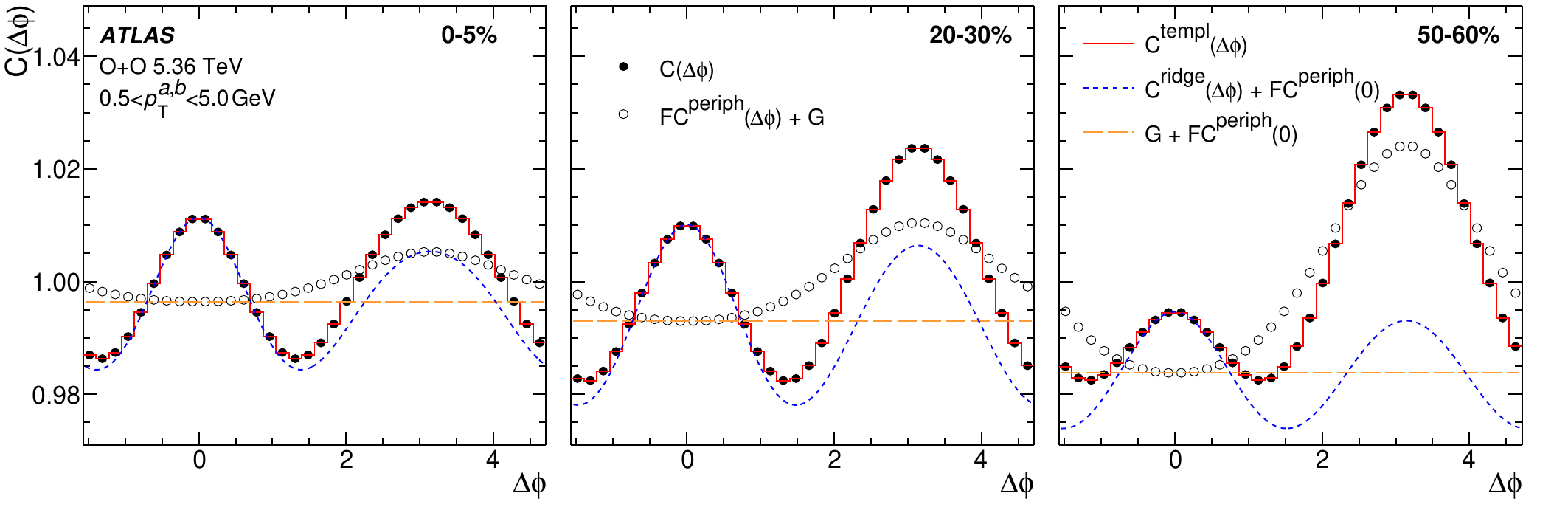}
\caption{
Template fits to correlation functions measured in \OO\ collisions
for the (left) 0--5\%, (center) 20--30\%, and (right) 50--60\%
centrality intervals.
The plots are for $0.5<\ptab<5$~\GeV.
The solid points indicate the measured \ctwophi,
and the continuous red line indicates the template fit, $C^{\mathrm{templ}}(\dphi)$.
The open points and dashed curves indicate the different components
of the template-fit, which are shifted along the $y$-axis by $G$
or by $FC^{\mathrm{periph}}(0)$,
where necessary, for presentation.
}
\label{fig:1d_2pc}
\end{figure}

Figure~\ref{fig:1d_2pc} shows examples of template fits for \OO\
collisions, where the template fit is denoted as
$C^{\mathrm{templ}}(\dphi)$.
In the measured correlations, the away-side peak is the largest in the
50--60\% centrality interval and decreases systematically in the
20--30\% and 0--5\% intervals.
A significant fraction of the away-side correlation is described by the
scaled peripheral reference ($F\cperi$ term in
Eq.~\ref{eq:tempfit_corr}).
The relative contribution of this term decreases monotonically toward
mid-central and central events, indicating that non-flow correlations
are most important in peripheral collisions.
The $C^\mathrm{ridge}(\Delta\phi)$ component of the template fit is
double-peaked in all intervals, reflecting the dominant contribution
from the $v_{2,2}$ term in Eq.~\ref{eq:tempfit_expansion}.

If the pair distribution is entirely determined by
a global single-particle distribution,
like in Eq.~\ref{eq:single}, the Fourier coefficients
of the \ctwophi\ (or $C ^\mathrm{ridge}(\Delta\phi)$)
factorize into the product of single-particle
anisotropies~\cite{Adare:2008ae} as
$\vnn(\pta,\ptb)=\vn(\pta)\vn(\ptb)$ and thus:
\begin{eqnarray}
\vn(\ptb)=\frac{\vnn(\pta,\ptb)}{\vn(\pta)}=\frac{\vnn(\pta,\ptb)}{\sqrt{\vnn(\pta,\pta)}},
\label{eq:factorization_2}
\end{eqnarray}
For all the \tpc\ results in this analysis, the \vn(\ptb) are
evaluated using Eq.~\ref{eq:factorization_2}
with $0.5<\pta<5~\GeV$.
The upper limit on \pta\ is chosen to suppress non-flow, which
increases at high \pt.
The template-fit method provides a more reliable treatment of non-flow
effects and is therefore regarded as the primary measurement method.
Results obtained with the \tpc\ method are also included in this
\papertype\ to illustrate the impact of non-flow removal achieved by
the template-fit procedure.

\subsection{Four-particle cumulants\label{sec:cumulant}}

Multi-particle cumulants extract higher‑order azimuthal correlations,
providing information about EbE flow fluctuations.
The cumulant measurements have the advantage of suppressing correlations from jets
and dijets, instead of relying on an explicit procedure to correct
$v_n$ as discussed in Section~\ref{sec:2pc}.
The cumulant of order $2k$, where $k$ is an integer,
involves correlations between $2k$ particles and suppresses
all correlations involving less than $2k$
particles, including non-flow correlations~\cite{Borghini:2001vi}.
The framework for the cumulant measurements is described in
Refs.~\cite{Bilandzic:2013kga,Bilandzic:2010jr,Jia:2017hbm},
but a concise description is provided here for completeness.

As mentioned before, the cumulant method involves the calculation of
$2k$-particle azimuthal correlations
$\langle \{2k\}_n \rangle$, and $2k$-particle cumulants, $c_n\{2k\}$,
for the $n^{\text{th}}$-order flow harmonics, where $k$ equals either one or
two in this \papertype.
The two- or four-particle azimuthal correlations in one event are
evaluated as~\cite{Bilandzic:2013kga,Bilandzic:2010jr,Jia:2017hbm}:
\begin{equation}
\langle \{2\}_n \rangle = \left\langle e^{in(\phi_1 - \phi_2)} \right\rangle
\label{eq:l2r_stan}
\end{equation}
\begin{equation}
\langle \{4\}_n \rangle = \left\langle e^{in(\phi_1 + \phi_2 - \phi_3 - \phi_4)} \right\rangle
\label{eq:l4r_stan}
\end{equation}
where ``$\langle \cdot \rangle$'' denotes a single-event average over
all pairs or quadruplets of distinct particles, respectively.
The averages from Eqs.~\ref{eq:l2r_stan} and \ref{eq:l4r_stan}
are expanded into products of per-particle normalized flow vectors~\cite{Voloshin:2007pc}:
\begin{equation}
\langle q_n \rangle = \left\langle e^{in(\phi)} \right\rangle,
\label{eq:q_vector}
\end{equation}
which provides an efficient calculation of multi-particle correlations.
The exact details of this procedure follow Ref.~\cite{HION-2017-02}.
These flow vectors are constructed with per-particle weights that
correct for detector non-uniformities, tracking inefficiency, and
contributions from fake tracks, similar to
Eq.~\ref{eq:effcorr} but with additional $\phi$-dependent
corrections applied.

In the ``standard'' cumulant method described so far, all $2k$-particle multiplets
involved in the calculations of $\langle \{2k\}_n \rangle$ are selected using the
entire detector acceptance.
To further suppress the non-flow correlations, which typically involve particles
emitted within a localized region in $\eta$, the particles can be grouped
into several ``subevents,'' each covering a non-overlapping $\eta$ interval~\cite{Jia:2017hbm}.
The multi-particle correlations are then constructed by correlating particles
between different subevents (two and three in this case), further reducing
non-flow correlations.
For two-subevent correlations, particles 1 and 2 in Eq.~\ref{eq:l2r_stan}
are selected from different regions of $\eta$.
Similarly, in Eq.~\ref{eq:l4r_stan},
the pair of particles 1 and 2 are selected from one
region of $\eta$, and the pair of particles 3 and 4 from another.
For the three-subevent correlations, permutations of different choices are made
and combined (see Ref.~\cite{HION-2017-02} for details).
For the results presented here, the subevents used for
the two-subevent method cover $-2.5<\eta<0$ and $0<\eta<2.5$,
and those for the three-subevent method cover
$-2.5<\eta<-2.5/3$, $-2.5/3<\eta<2.5/3$ and $2.5/3<\eta<2.5$.

The two- and four-particle cumulants are then obtained from the
azimuthal correlations as:
\begin{align}
c_n\{2\} &= \langle\!\langle \{2\}_n \rangle\!\rangle, \label{eq:c2_stan} \\
c_n\{4\} &= \langle\!\langle \{4\}_n \rangle\!\rangle - 2\langle\!\langle \{2\}_n \rangle\!\rangle^2, \label{eq:c4_stan}
\end{align}
where ``$\langle\!\langle \cdot \rangle\!\rangle$'' represents
the average of $\langle \{2k\}_n \rangle$ over an event ensemble.
In the absence of non-flow correlations, $c_n\{2k\}$ reflects
the moments of the distribution of the flow coefficient $v_n$:
\begin{align}
c_n\{2\} & = \langle v_n^2 \rangle  , \label{eq:vn_stan} \\
c_n\{4\} &= \langle v_n^4 \rangle - 2\langle v_n^2 \rangle^2  \,.\nonumber
\end{align}

The \vn\ measured by the 2-particle cumulants is defined
as $\vntwo=\sqrt{c_n\{2\}}$.
In the absence of non-flow effects, the \vn\ measured by the 2-particle
cumulant is identical to that measured using the \tpc\ method.
However, non-flow contributions lead to differences between the two.
Henceforth, the \vn\ measured with the \tpc\ and template-fit,
methods are denoted as \vntpc\ and \vntf, respectively.
When making general statements without reference to a specific method,
$\vn$ obtained from any two-particle correlation technique
is denoted simply as \vntwo.

Under the Gaussian model of eccentricity fluctuations~\cite{Voloshin:2007pc},
the 2- and 4-particle cumulants can be expressed in terms of an average
geometry-driven component $\bar{x}_n$ and a fluctuation component
$\Delta_n$ as:
\begin{align}
c_n\{4\} = -\bar{x}_n^4, \quad c_n\{2\} \equiv \langle v_n^2 \rangle  = \bar{x}_n^2 + \Delta_n^2.
\label{eq:cnFluc}
\end{align}
This implies $c_n\{4\} < 0$, and a positive $c_n\{4\}$ signals onset
of non-Gaussian flow fluctuations, or significant non-flow contamination.
If the sign constraints are obeyed, the corresponding four-particle
flow coefficient is then defined as:
\begin{align}
v_n\{4\} = \sqrt[4]{-c_n\{4\}},
\end{align}
and measures the contribution to the flow from the
average-geometry component only.
A comparison of the results of the standard and the two- and three-subevent
cumulants in the \OO\ and \NeNe\ collisions shows that the
two- and three-subevent results are consistent.
This demonstrates that both subevent methods effectively suppress
non-flow in the measured phase space region.
For this reason, the \vnfour\ results presented in
this \papertype\ uses the two-subevent method.

\section{Systematic uncertainties}
\label{sec:systematics}

The systematic uncertainties of the measured \vn\ using the \tpc,
template-fit, and multi-particle cumulant methods are described in
this section.
The following sources of systematic uncertainty are considered:

\begin{table}
\caption{
The contributions to the systematic uncertainty of
\vn\ in \OO\ collisions from different sources,
as function of centrality.
The contributions are expressed in percentages.
Items 1--5 are common to all the methods used here (\tpc, template-fit and cumulants).
Item 6 is specific to the 2PC and template-fit methods.
Item 7 is specific to the template-fit method.
Item 8 is specific to the cumulant method.
The uncertainties are shown for the integrated \pt\ interval of 0.5--5~\GeV.
\label{tab:uncSP}
}
\begin{center}
\begin{tabular}{|l c|c|c|}
\hline
Source   &  harmonic order  & 0--40\% & 40--70\%   \\
&       & [\%] &[\%]       \\
\hline
1. MC closure            & \vtwo--\vfour       & 1    & 1  \\
\hline
\multirow{3}{*}{2. Track selection }
& \vtwo                 &  0.5   &  0.5      \\
& \vthree               &  0.75  &  0.75     \\
& \vfour                &  1.5   &  1.5      \\
\hline
3. Tracking efficiency  & \vtwo--\vfour       & 0.25   & 0.25   \\
\hline
\multirow{3}{*}{4. Centrality definition}
& \vtwo       &   0.2       & 0.2--0.6    \\
& \vthree     &  0.2--1.0   & 1--2        \\
& \vfour      &   0.2       & 0.2--0.6    \\
\hline
5. Residual pileup       & \vtwo--\vfour       & 0.2    & 0.2    \\

\hline
\multirow{3}{*}{6. Event-mixing}
& \vtwo                 &  0.25&  0.25     \\
& \vthree               &  0.5 &  0.5      \\
& \vfour                &  1   &  1        \\
\hline
\multirow{3}{*}{7. Peripheral reference}
& \vtwo                 &  0.5       &     0.5--3  \\
& \vthree               &  0.75--3.5 &     3.5--12  \\
& \vfour                &  1.0--4.5  &     4.5--20 \\
\hline

8. Flattening procedure    & \vtwo--\vthree      &  0.25 & 0.25    \\
\hline
\end{tabular}
\end{center}
\end{table}

\begin{enumerate}
\item \textbf{MC closure:}
The MC closure test compares $v_{n}^{\mathrm{gen}}$,
obtained from MC generated particles, with
$v_{n}^{\mathrm{reco}}$, obtained by applying the full analysis
procedure to reconstructed tracks in the MC simulation, as done in
the data analysis.
The difference between the two is within 1\% across the \pt\ and
multiplicity ranges considered in this analysis and is conservatively
assigned as a systematic uncertainty.
This uncertainty accounts for residual reconstruction effects not
corrected in the data analysis.

\item \textbf{Track selection:}
The track selection criteria control the relative contributions of
genuine charged particles and fake tracks entering the analysis.
The stability of the results with respect to the track selections is
evaluated by varying the requirements applied to reconstructed tracks
and including the resulting variation in \vn\ as a systematic
uncertainty.
The results obtained with the nominal selections are compared with those
using the tighter criteria described in Section~\ref{sec:detector}.
The differences depend on the harmonic order and are between 0.5--1.5\%.

\item \textbf{Tracking efficiency:}
The uncertainty in the reconstruction efficiency and fake-rate due to
ID material modeling in the \GEANT\ simulation is accounted for by
evaluating the efficiency and fake-rates in alternate MC samples.
In each sample, a single modification is applied to the ATLAS ID
geometry: the passive material of the ID is increased by 5\%, or the
passive material of the IBL by 10\%, or the passive material in the
services region by 25\%.
These variations capture the full range of data–MC differences observed
in dedicated studies of the ID material~\cite{PERF-2015-07}.
The variation in the results when using these alternative models is
taken as a systematic uncertainty.
This uncertainty is less than $0.25\%$.

\item \textbf{Centrality definition:}
The centrality definitions used to classify the events into centrality
percentiles have a $\sim$1\% (2\%) uncertainty associated with
them in the \OO\ (\NeNe) measurements.
This arises from uncertainties in the fraction of the
inelastic \OO\ and \NeNe\ cross-sections accepted by the triggers used
in this analysis and is estimated from the Glauber fits to the
\fcalsum\ distributions~\cite{HION-2011-05,HION-2016-06}.
The impact of this uncertainty on the \vn\ is evaluated by varying the
\fcalsum\ thresholds that define the centrality intervals,
re-evaluating the \vn, and assigning the observed variation as a
systematic uncertainty.
This uncertainty is negligible in most central collisions, and increases
systematically for more peripheral collisions.

\item \textbf{Residual pileup:}
Pileup events dilute the measured \vn\ as there are no correlations
between independent collisions.
The estimated residual pileup is at most 0.2\%  in any centrality or
multiplicity interval considered in this \papertype.
Because the maximum possible dilution cannot exceed the pileup rate,
the entire residual pileup rate of 0.2\% is conservatively assigned
as a systematic uncertainty on the \vn.

\item \textbf{Event-mixing:}
As mentioned before, the \tpc\ method uses event-mixing
to account for detector acceptance effects.
The nominal mixing matches events that are within a
\zvtx\ separation of less than 20~mm.
Alternate mixing criteria, where the matching is restricted
to within 10~mm and relaxed to 200~mm are used, and
the maximum variation in the results is included as a
systematic uncertainty.
This uncertainty is of order 1\% and only affects the \tpc\
and template-fit measurements.

\item \textbf{Peripheral reference:}
For the centrality-dependent measurements, the nominal template-fit
procedure uses events more peripheral than 80\% for building the
peripheral reference \cperi.
For the multiplicity-dependent results, the peripheral references
are  built from events from the same collision system that have \nchrec\ less than 20.
For evaluating uncertainties associated with the assumptions made in the
template fit analysis, the centrality and multiplicity dependence measurements
are repeated with the peripheral
reference built from 5.02~\TeV\ \pp\ events with \nchrec\ less than 20,
as was done in Ref.~\cite{HION-2019-03}.
The difference between the results with this alternate choice of peripheral
reference are included as a systematic uncertainty.
This uncertainty is relevant only for the template-fit method.
This uncertainty is the dominant contribution in peripheral centralities.
In this region, the measured correlation, \ctwophi, becomes very
similar in shape to the peripheral reference correlation, \cperi, used in Eq. 4.
As a result, the template decomposition becomes increasingly sensitive
to small differences between the two, leading to a large relative uncertainty
in the extracted \vn.

\item \textbf{Flattening procedure:}
The cumulant measurements use a flattening procedure to remove
detector non-uniformities in $\phi$ ~\cite{HION-2017-02}.
As a conservative estimate of the systematic related to the
flattening procedure, the measurements were repeated with the
flattening removed, and the resulting variations of $\sim0.25\%$ are included
as a systematic uncertainty on the \vnfour.
\end{enumerate}

Table~\ref{tab:uncSP} summarizes the final systematic uncertainties for the
integrated \pt\ interval of 0.5--5.0~\GeV\ in the \OO\ measurements.
Except for the uncertainties related to the peripheral reference and
centrality definition, all other uncertainties are conservatively
taken to be constant and sufficiently large to cover the variations
across all centrality (or multiplicity) and \pt\ intervals studied here.
The dominant uncertainties are the uncertainties related to the
MC-closure and the peripheral reference variation.
For \NeNe, the assigned uncertainties are taken to be the same
as for \OO, with values chosen large enough to cover both systems, except for the centrality-definition uncertainty,
which is about twice as large.
These individual uncertainties are added in quadrature to obtain
total uncertainties.
For ratios of the \NeNe\ to \OO\ \vn\ measurements, all sources of
systematic uncertainty are treated as correlated,
with the exception of the residual pileup.
The latter is not considered correlated,
since pileup rates can differ between the two systems due to their
distinct running conditions.
The remaining systematic uncertainties are treated as correlated between the
two collision systems.
The charged-particle multiplicities in \NeNe\ are only about 20\% higher
than those in \OO, leading to very similar detector occupancy and
reconstruction conditions.
Consequently, the systematic effects related to tracking and detector acceptance,
namely the MC closure, track-selection, tracking-efficiency, event-mixing and flattening
uncertainties, are expected to be identical in both systems.
The peripheral reference related uncertainty is evaluated by
using an identical low-multiplicity \pp\ reference for both systems,
and is also treated as a correlated uncertainty.
In addition, the centrality-definition uncertainty is treated as
correlated, as it is dominated by the modeling of nucleon distributions
in the Glauber calculation -- such as the choice between hard-sphere and diffuse nucleons,
or the inclusion of nucleon clustering -- which affects both collision systems in the same way.

\section{Results}
\label{sec:result}

\begin{figure}
\centering
\includegraphics[width=0.99\linewidth]{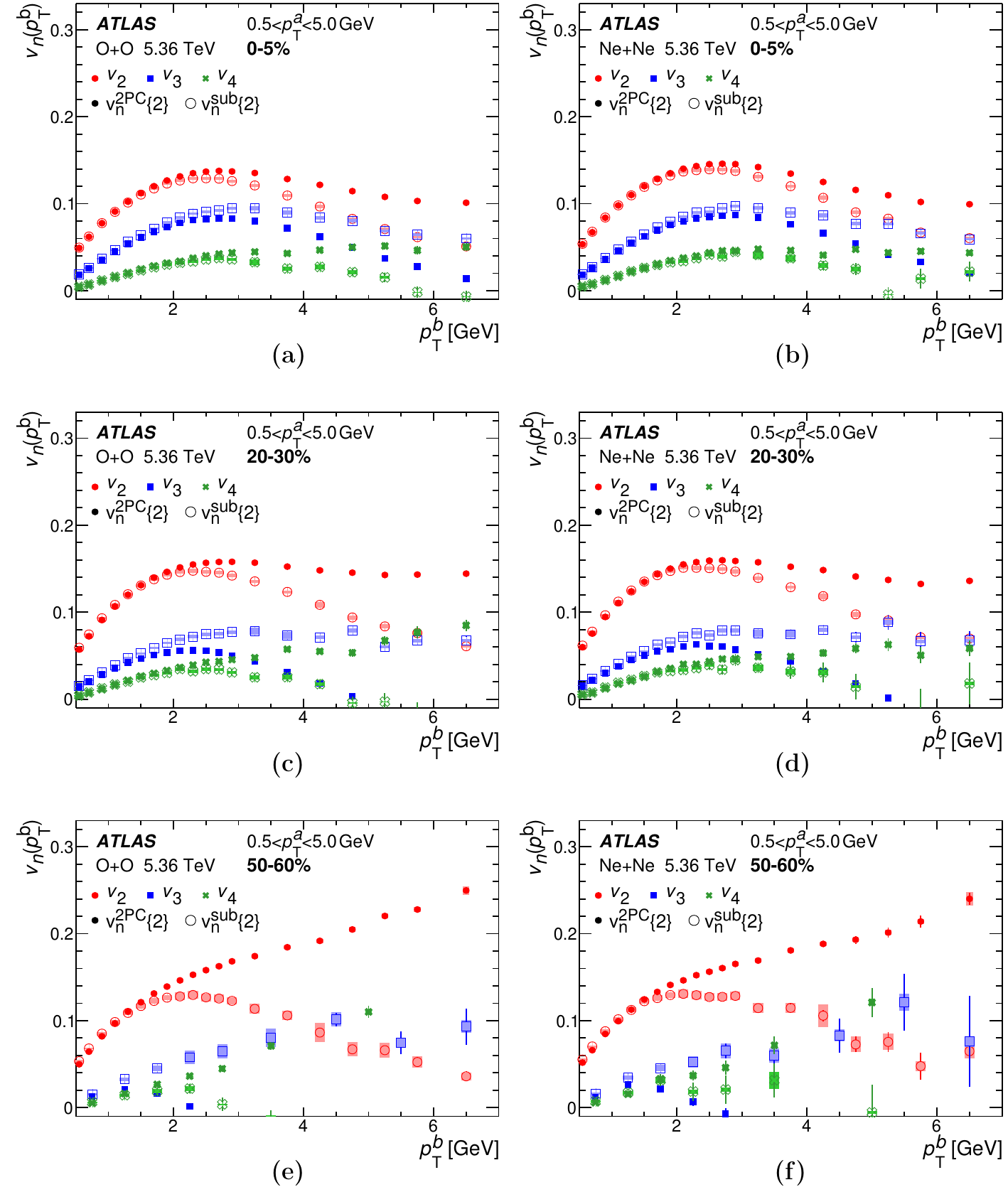}
\caption{
The $\ptb$ dependence of the \vntpc\ and \vntf\ in
the (a,~c,~e) \OO\ and (b,~d,~f) \NeNe\ collisions,
for the (a,~b) 0--5\%, (c,~d) 20--30\% and (e,~f) 50--60\% centrality intervals.
The solid and open points show the results obtained using
the \tpc\ and the template-fit methods, respectively.
The vertical lines and vertical bars indicate statistical and
systematic uncertainties, respectively.
}
\label{fig:ptDep_oo}
\end{figure}

Figure~\ref{fig:ptDep_oo} shows the measured \vntf\
and the \vntpc\ as a function
of \ptb\ in \OO\ and \NeNe\ collisions at $\snn=5.36$~\TeV.
Comparing the template fit \vn\ and the \tpc\ \vn,
a large potential non-flow contribution is observed in
the \tpc\ results, particularly at higher \ptb\ and
in peripheral collisions.
The \vtwo\ results exhibit
a linear rise at low \ptb, followed by a decrease
in the range of $2$--$4$~\GeV\ for the 0--5\% and 20--30\%
centrality intervals.
This behavior is qualitatively similar to what is observed
in other collision systems, from \PbPb\ to \pp~\cite{HION-2016-01, HION-2019-03,HION-2015-09,HION-2016-06}.
A hierarchy is also observed, with the magnitude of \vn\
decreasing as $n$ increases, in agreement with observations
from other systems.
In peripheral events (50--60\% centrality), or at \pt\
above approximately $3~\GeV$,
significant differences are observed between the \tpc\ and
template-fit measurements, highlighting the impact of
non-flow background correlations on the \tpc\ method.

\begin{figure}
\centering
\includegraphics[width=0.99\linewidth]{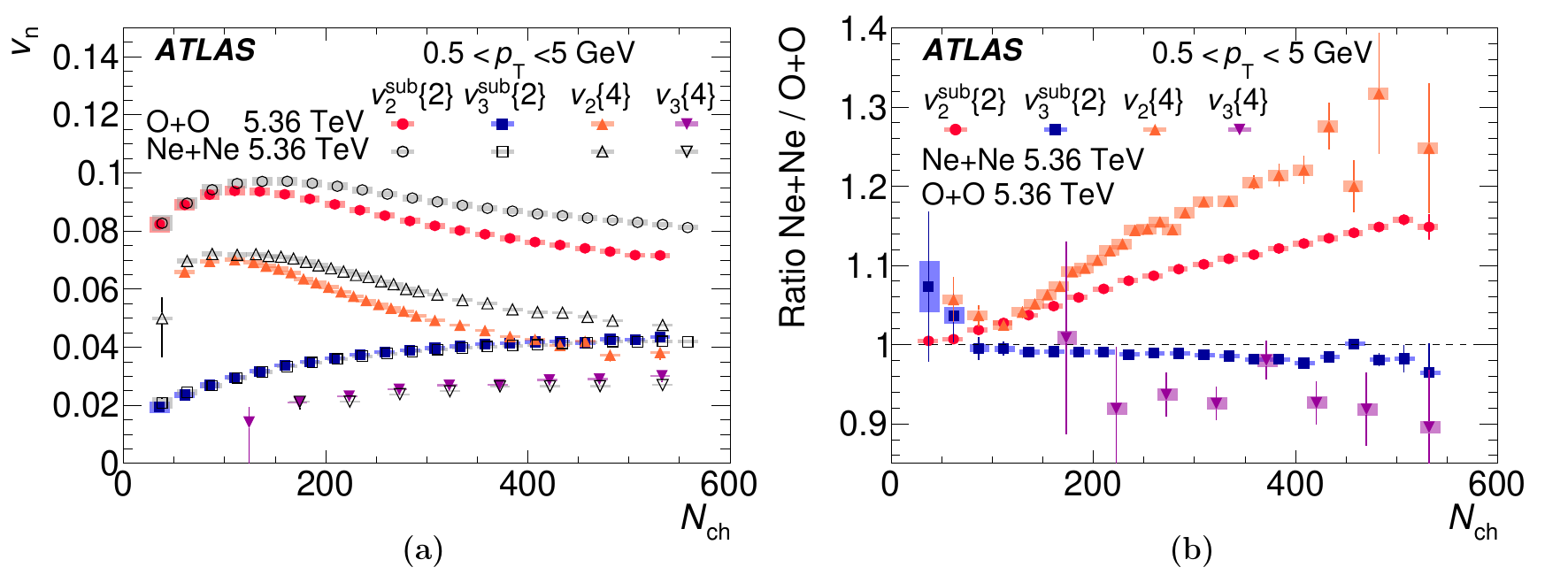}
\caption{
(a) Flow harmonics $\vntf$ and $\vnfour$ for the elliptic
and triangular moments in \OO\ collisions
(colored solid markers) and \NeNe\ collisions
(open black markers) at $\snn=5.36$~\TeV,
shown as a function of $N_\text{ch}$.
The \NeNe\ points are slightly displaced along the
horizontal axis for visual clarity.
(b) Ratios of the flow coefficients in \NeNe\ to those
in \OO\ collisions, corresponding to the results
in panel (a).
Ratios are shown only for the 0--50\% centrality range to suppress
large statistical fluctuations in more peripheral intervals.
The vertical lines and vertical bars indicate statistical and
systematic uncertainties, respectively.
For the ratios, systematic uncertainties correlated between
the \OO\ and \NeNe\ measurements largely cancel.
}
\label{fig:multDep}
\end{figure}

The hydrodynamic response to the initial geometry is best
studied as a function of variables that are directly
impacted by it, chiefly charged-particle multiplicity
and centrality.
Figure~\ref{fig:multDep}a shows \vntf\
and \vnfour\ for charged particles with $0.5<\pt<5.0$~\GeV\
in \OO\ and \NeNe\ collisions as a function of $N_{\mathrm{ch}}$.
In both systems,  \vntf\ is observed to exceed \vnfour\
for the same harmonic order $n$, which reflects the
positive contribution of fluctuations to \vntwo\ as
described in Eq.~\ref{eq:cnFluc}.
The difference between the  $v_2\{2\}$ and $v_2\{4\}$
increases with increasing multiplicity indicating
the increased role of the fluctuation component,
$\Delta_{2}^{2}$ in Eq.~\ref{eq:cnFluc},
relative to the mean geometry component, $\bar{x}_{2}^{2}$,
going from peripheral to central collisions.

To compare \OO\ and \NeNe\ \vn\ directly, their ratio is plotted in
Figure~\ref{fig:multDep}b.
For the elliptic flow, $v_{2}$, there is a growing enhancement in
\NeNe\ collisions with increasing $N_\mathrm{ch}$.
These can be attributed to  centrality-dependent changes in the overlap
geometry and neon’s deformed nuclear structure, which will be further
discussed below, in the context of the centrality-dependent results.
Conversely, there is a striking similarity of $\vntf[3]$ in \OO\ and \NeNe\
collisions as a function of $N_\text{ch}$.
This provides evidence that the hydrodynamic response and the fluctuation-driven
component of the initial-state geometry depend primarily on the
particle density.
The residual difference and systematically smaller $v_3\{4\}$ may be an
indication of differing average triangularity in the initial-state geometry.

\begin{figure}
\centering
\includegraphics[width=0.99\linewidth]{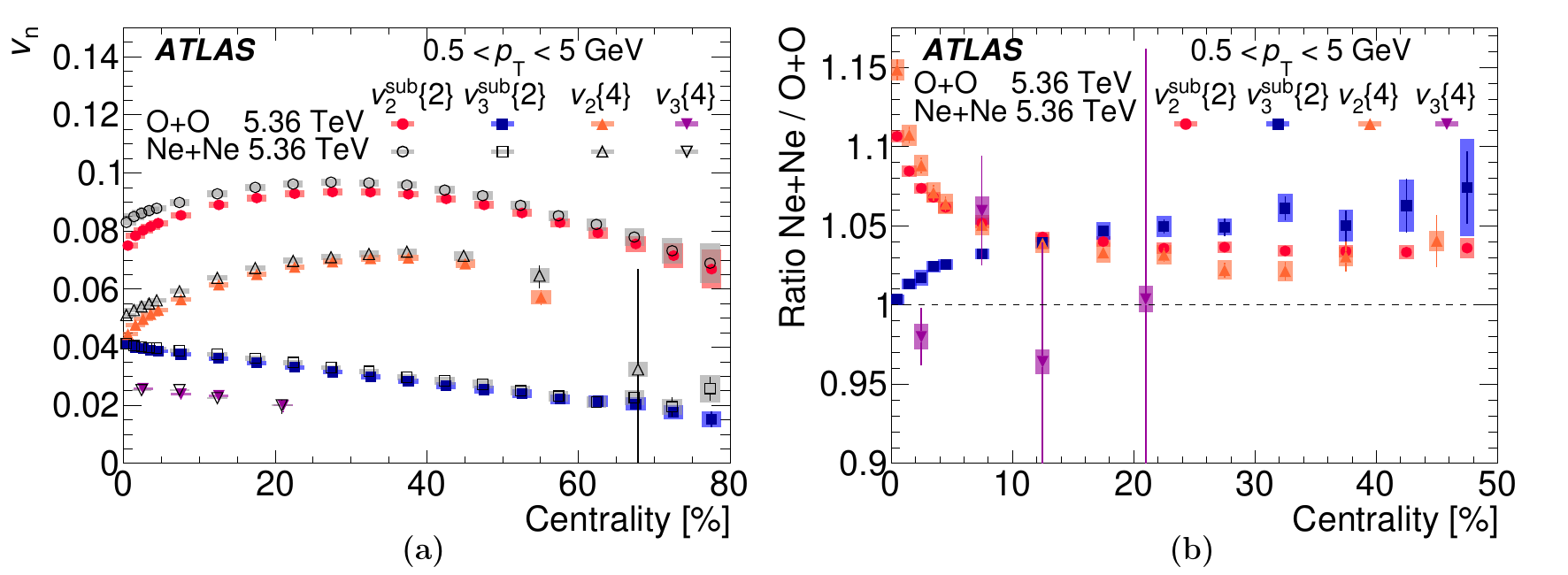}
\caption{
(a) Flow harmonics $\vntf$ and $\vnfour$ for the elliptic
and triangular moments in \OO\ collisions
(colored solid markers) and \NeNe\ collisions
(open black markers) at $\snn=5.36$~\TeV,
shown as a function of centrality.
The \NeNe\ points are slightly displaced along the
horizontal axis for visual clarity.
(b) Ratios of the flow coefficients in \NeNe\ to those
in \OO\ collisions, corresponding to the results
in panel (a).
The vertical lines and vertical bars indicate statistical and
systematic uncertainties, respectively.
For the ratios, systematic uncertainties correlated between
the \OO\ and \NeNe\ measurements largely cancel.
}
\label{fig:centDepl}
\end{figure}

To compare the geometric aspects of oxygen and neon nuclei,
Figure~\ref{fig:centDepl}a presents the results
as a function of collision centrality, using the same
kinematic requirements and analysis techniques as
in Figure~\ref{fig:multDep}.
Although light-ion collisions have large initial-state geometry fluctuations
and possibly nuclear deformations, a large contribution to the
elliptic geometry is the event-averaged lenticular shape.
This event-averaged shape depends on the ratio of the impact parameter
to the nuclear radius, which is highly correlated with centrality.
For this reason there is closer agreement observed between \OO\ and \NeNe\
as a function of centrality than $N_\text{ch}$.
Thus, the ratio of \NeNe\ to \OO\ flow harmonics is calculated
as a function of centrality, in an attempt to remove these
average geometric effects.
This ratio is shown in
Figure~\ref{fig:centDepl}b.
For both \vntf[2]\ and $v_{2}\{4\}$, a relatively centrality-independent
ratio is observed for centralities more peripheral than 10\%,
as opposed to the ratios as a function of
$N_\mathrm{ch}$ (Figure~\ref{fig:multDep}b).
However, over the 0–10\% centrality interval, a growing enhancement appears
toward the most central collisions.
This enhancement is more pronounced in the four-particle cumulant measurement.
A similar behavior is observed for the triangular moment: the \vntf[3]\ ratio
exhibits a sudden decrease in the 0--10\% centrality interval.
For the $v_3\{4\}$ ratio, the statistical uncertainties are too large to observe
clear trends.

\begin{figure}
\centering
\includegraphics[width=0.99\linewidth]{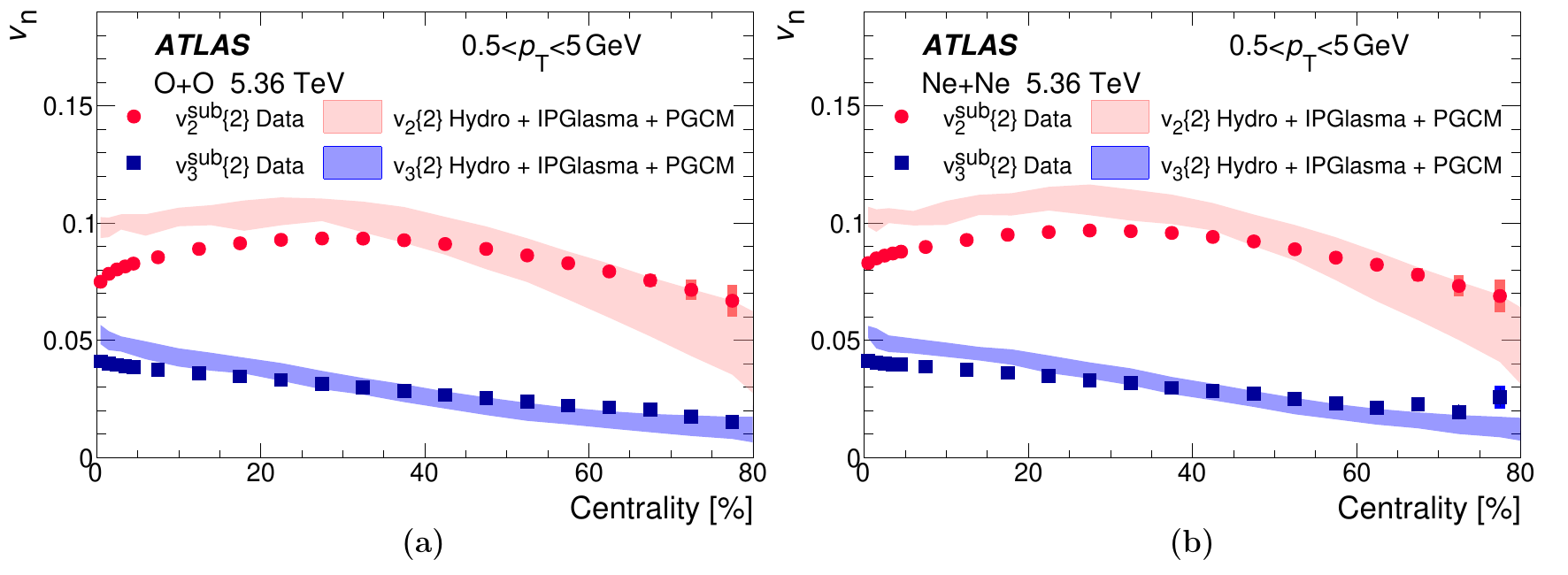}
\caption{
Comparison of $\vntf[2]$ and $\vntf[3]$ with theory calculations
from the model described in Ref.~\cite{Mantysaari:2025tcg},
for (a) \OO\ and (b) \NeNe\ collisions.
The theory calculations are labeled as ``Hydro$+$IPGlasma$+$PGCM.''
For the data, the vertical lines and vertical bars indicate statistical and
systematic uncertainties, respectively.
The data uncertainties are sometimes too small to be visible.
The bands for the theory calculations represent the combined
statistical and systematic uncertainties.
\label{fig:theory_OO_compare_centdep_aux1}
}
\end{figure}

Figure~\ref{fig:theory_OO_compare_centdep_aux1} shows the comparisons of
the measured \vntf\ with the template-fit method to \vntwo\ calculations
from the model described in Ref.~\cite{Mantysaari:2025tcg}.
The model combines the Projected Generator Coordinate Method (PGCM),
an \textit{ab initio} nuclear structure model for oxygen and neon,
with the IPGlasma framework, including JIMWLK
evolution~\cite{Schenke:2012wb}, to calculate the initial energy
densities of the colliding nuclei.
These energy densities are then evolved using viscous relativistic
hydrodynamics (MUSIC~\cite{Schenke:2010nt}).
Finally, a hadronic afterburner (UrQMD~\cite{Bleicher:1999xi,Bass:1998ca})
is applied to obtain the \vn.
The centrality is determined with the charged-particle multiplicity
at mid-rapidity.
The model is denoted as ``Hydro$+$IPGlasma$+$PGCM'' in the figure.
The calculations overpredict the \vtwo\ and \vthree\ in central collisions,
but are consistent within uncertainties with the measurements in more
peripheral collisions.
These comparisons indicate that the light-ion measurements can provide
additional constraints for tuning existing theory models.

\begin{figure}
\centering
\includegraphics[width=0.99\linewidth]{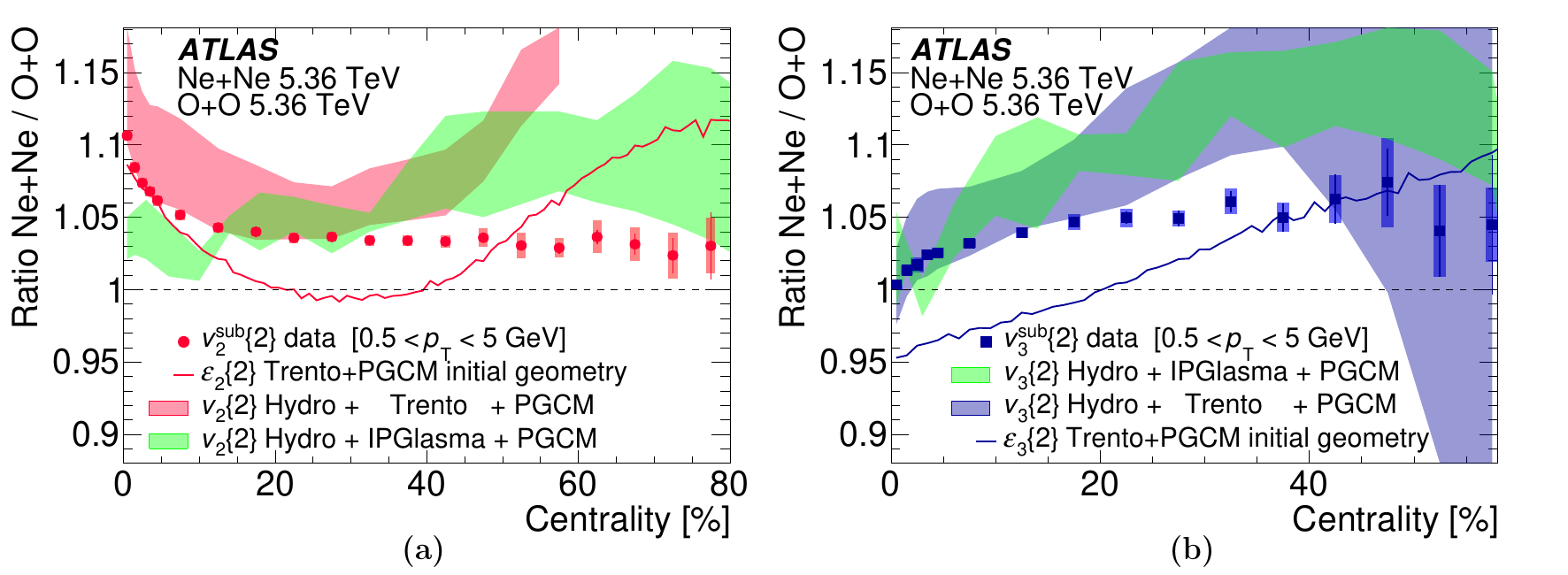}
\caption{
Comparison of the measurements to theory predictions for the
ratio of (a) $v_2\{2\}$ and (b) $v_3\{2\}$
between \NeNe\ and \OO.
The \vntwo\ in the data are obtained from the template fit method.
The Hydro$+$Trento$+$PGCM theory is taken from
Ref.~\cite{Giacalone:2024luz} ($0.3<\pt<3.0$~GeV) and the Hydro$+$IPGlasma$+$PGCM
is an extension of Ref.~\cite{Mantysaari:2025tcg} ($0.5<\pt<5.0$~GeV).
The ratios are also shown for the eccentricities, a quantification of the elliptic ($\varepsilon_2$) or triangular ($\varepsilon_3$) shape of the initial-state energy density, using the Trento$+$PGCM model.
For the data, the vertical lines and vertical bars indicate statistical
and systematic uncertainties, respectively.
The bands for the theory calculations represent combined
statistical and systematic uncertainties.
}
\label{fig:theoryRatio}
\end{figure}

Figure~\ref{fig:theoryRatio} shows comparisons of the measured
ratios for \vntwo\ between \NeNe\ and \OO\ collisions,
to two theory calculations.
The first is the Hydro$+$IPGlasma$+$PGCM model introduced above.
The second model is described in Ref.~\cite{Giacalone:2024luz}
and uses PGCM~\cite{Lee:2008fa} to
determine the nuclear configurations of both oxygen and neon,  which
are then collided, generating  initial-state energy densities with
the Trento model~\cite{Moreland:2014oya}.
These events are then simulated with the Trajectum hydrodynamic
framework~\cite{Nijs:2020roc,Nijs:2021clz}.
This model is labeled as ``Hydro$+$Trento$+$PGCM.'' %
The model uses mid-rapidity multiplicity to determine centrality.
The fluctuations in the model calculations are statistical in nature.
The measured \vntwo\ are obtained from the template-fit method.
The event-averaged eccentricity, a quantification of the magnitude of ellipticity
$\epsilon_2$ and the triangularity $\epsilon_3$, ratios for a particular Trento
parameter set~\cite{Giacalone:2023cet} are also shown in Figure~\ref{fig:theoryRatio}.
The data and Hydro$+$Trento$+$PGCM model agree quantitatively,
at the edge of the theory uncertainties, in $v_2\{2\}$ in central collisions.
The large enhancement in the $v_2\{2\}$ ratio in the 0--10\% central collisions
observed in the hydrodynamic theory calculation is attributed to the
elongated shape of the neon nucleus, leading to an elliptic initial-state,
as can be seen in the $\epsilon_2\{2\}$ ratio.
The qualitative behavior of the measured $v_2\{2\}$ ratio
supports this picture.
The Hydro$+$IPGlasma$+$PGCM calculation does not capture this proposed
geometrically driven behavior in the 0--10\% centrality range.
This inability to capture geometric effects in the latter calculation,
is also reflected in the data-theory disagreement in
Figure~\ref{fig:theory_OO_compare_centdep_aux1}.
In both \NeNe\ and \OO\ collisions, the measured \vtwo\ decreases in
the most central collisions, but this trend is not predicted by the model.
Because the nuclear structure model PGCM can capture geometric effects,
as seen in the Hydro$+$Trento$+$PGCM \vtwo\ ratio agreement with the data,
the disagreement with the Hydro$+$IPGlasma$+$PGCM calculations could arise
in the IPGlasma-with-JIMWLK portion of the initial-state model.
It has been noted~\cite{Mantysaari:2025tcg} that this aspect of the
model creates large multiplicity fluctuations, reducing the
correlation between the centrality and impact parameter.
This can lead to the inability of the model to reproduce
the centrality dependence trends observed in
Figures~\ref{fig:theory_OO_compare_centdep_aux1} and~\ref{fig:theoryRatio}.
For both model comparisons presented in Figure~\ref{fig:theoryRatio},
the experimental uncertainties are significantly smaller than the
statistical and systematic uncertainties associated with the
model calculations.
Consequently, further refinement of the theory predictions is
required to fully capitalize on the precision of the data.

\begin{figure}
\centering
\includegraphics[width=0.99\linewidth]{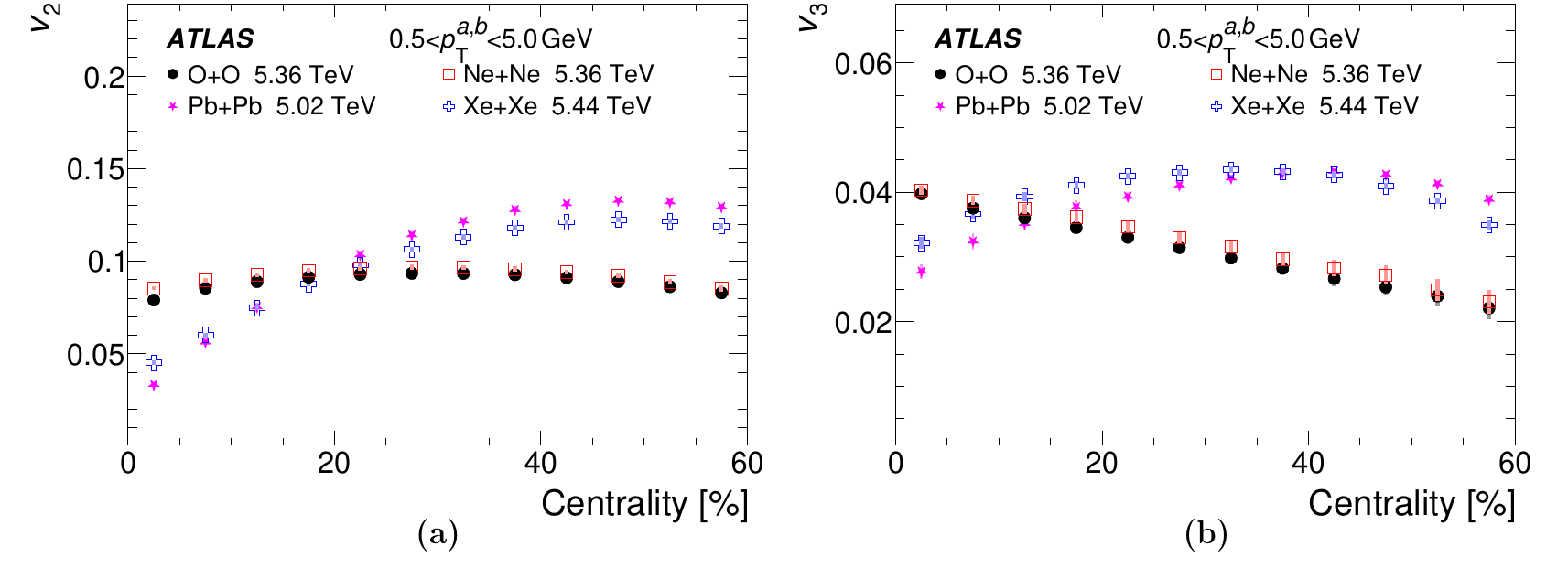}
\caption{
Comparisons of the (a) \vtwo\ and (b) \vthree\ measured by the
template-fit method in
\OO\ and \NeNe\ collisions with prior measurements in
\XeXe\ and \PbPb\  collisions from Ref.~\cite{HION-2019-03},
as a function of centrality.
The \XeXe\ and \PbPb\ results are obtained with the \tpc\ method.
The plots are for $0.5<\ptab<5$~\GeV.
The vertical lines and vertical bars indicate statistical and
systematic uncertainties, respectively.
\label{fig:vn_compare_XePb_cent}
}
\end{figure}

\begin{figure}[h!]
\centering
\includegraphics[width=0.99\linewidth]{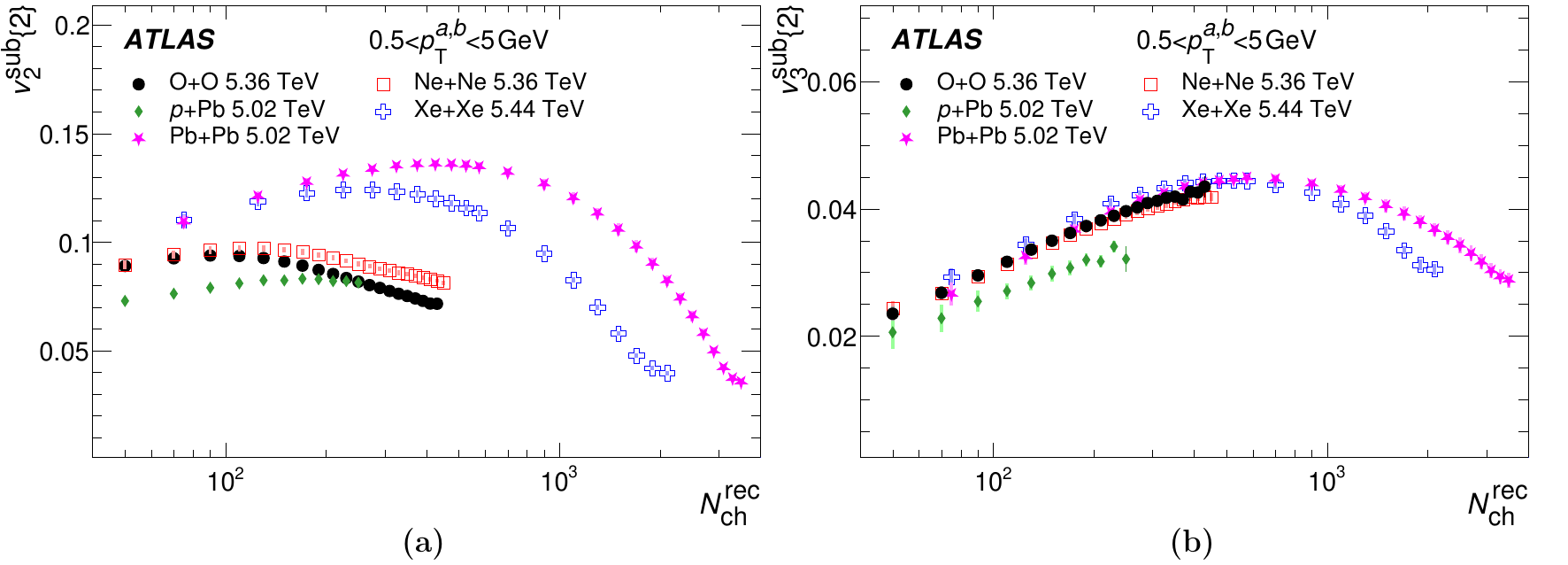}
\caption{
Comparison of the (a) \vntf[2]\ and the (b) \vntf[3]\  measured in
\OO\ and \NeNe\ collisions with prior measurements in
\pPb, \XeXe\ and \PbPb\ collisions from Ref.~\cite{HION-2019-03},
as a function of the reconstructed charged-particle
multiplicity (\nchrec).
The vertical lines and vertical bars indicate statistical and
systematic uncertainties, respectively,
and are often too small to be visible.
}
\label{fig:vn_compare_XePb_mult}
\end{figure}

Figure~\ref{fig:vn_compare_XePb_cent} shows comparisons of \vntf\
measurements in \OO\ and \NeNe\ collisions to previous \vntpc\
results in \XeXe\ collisions at 5.44~\TeV\ and \PbPb\ collisions at
5.02~\TeV\ from Ref.~\cite{HION-2019-03}, as a function of
centrality.
In heavy-ion collisions, the \vtwo\ is largely determined by the
elliptic geometry of the nuclear overlap, which is strongly correlated
with centrality and thus drives the pronounced centrality dependence
of \vtwo.
By contrast, the \vtwo\ in light-ion systems shows a much weaker
centrality dependence, highlighting the dominant role of EbE geometry
fluctuations in these systems.
The \vthree\ in \XeXe\ and \PbPb\ collisions exhibits a non-monotonic
centrality dependence, increasing from central to mid-central
collisions and then decreasing.
In light-ion collisions, however, the \vthree\ decreases monotonically
from central to peripheral events.
Figure~\ref{fig:vn_compare_XePb_mult} shows similar comparisons of
\vntf\ as a function of reconstructed charged-particle multiplicity,
\nchrec, including results from \pPb\ collisions at 5.02~\TeV.
The most notable feature is the very similar magnitude and \nchrec\
dependence of \vthree\ in both light- and heavy-ion collisions.
This stands in stark contrast to the centrality dependence of \vthree,
which differs strongly between light- and heavy-ion systems.
Interestingly, the \pPb\ \vthree\ is also comparable in magnitude to
the light- and heavy-ion results at the same multiplicities and shows
a qualitatively similar multiplicity dependence.

\FloatBarrier
\section{Conclusion}
\label{sec:conclusion}

This paper presents the first comprehensive measurements of anisotropic
flow coefficients $v_{n}$ ($n=2$–4) in $\snn=5.36$~\TeV\ \OO\ and
\NeNe\ collisions with the ATLAS detector at the LHC.
The results are obtained using two-particle (template-fit) and
four-particle (subevent cumulant) methods, and explore the dependence
of $v_{n}$ on transverse momentum, multiplicity, and centrality in
each system.
To isolate the role of initial-state geometry, ratios of \NeNe\ to \OO\
$v_{n}$ are also studied.

The measurements reveal a characteristic rise and fall of $v_{2}$ with
\pt, peaking at $2$--$4$~\GeV, large event-by-event fluctuations in the
most central collisions, and a clear hierarchy $v_{2}>v_{3}>v_{4}$,
consistent with hydrodynamic expectations.
Multiplicity- and centrality-dependent ratios of $v_{2}\{2\}$ and
$v_{2}\{4\}$, taken between \NeNe\ and \OO\ collisions, reveal a
marked enhancement in \NeNe, consistent with the elongated nuclear
shape of neon and reproduced by model calculations.
For \vtwo, the four-particle cumulant ratios provide the greatest
sensitivity to these geometric effects.
For $v_{3}\{2\}$, the \NeNe\ values exceed those in \OO\ for most
centralities, except in the 0--1\% most central events where they are
comparable.
Detailed comparisons of these measurements with model predictions may
also provide new sensitivity to $\alpha$-clustering in the oxygen
nucleus~\cite{YuanyuanWang:2024sgp}.
Compared to heavy-ion collisions, the \vtwo\ in light-ions shows a much
weaker centrality dependence, underscoring the enhanced role of EbE
geometry fluctuations in these systems.
As a function of event multiplicity, the $v_{3}$ values in \OO\ and
\NeNe\ are quantitatively similar to those measured in \PbPb, \XeXe,
and \pPb\ collisions.

These light-ion results offer stringent constraints on hydrodynamic
models incorporating \textit{ab initio} nuclear-structure inputs, and
provide unique insight into how nucleon-level geometry influences
collective flow in small QGP droplets.
Future comparisons with detailed model calculations could further
constrain key medium and geometric properties, including the
temperature dependence of $\eta/s$, the role of the pre-hydrodynamic
phase, mechanisms of energy deposition in hadronic collisions, and
other initial-state effects.

\section*{Acknowledgements}
%

%

%
%

%
%

We thank CERN for the very successful operation of the LHC and its injectors, as well as the support staff at
CERN and at our institutions worldwide without whom ATLAS could not be operated efficiently.

The crucial computing support from all WLCG partners is acknowledged gratefully, in particular from CERN, the ATLAS Tier-1 facilities at TRIUMF/SFU (Canada), NDGF (Denmark, Norway, Sweden), CC-IN2P3 (France), KIT/GridKA (Germany), INFN-CNAF (Italy), NL-T1 (Netherlands), PIC (Spain), RAL (UK) and BNL (USA), the Tier-2 facilities worldwide and large non-WLCG resource providers. Major contributors of computing resources are listed in Ref.~\cite{ATL-SOFT-PUB-2025-001}.

We gratefully acknowledge the support of ANPCyT, Argentina; YerPhI, Armenia; ARC, Australia; BMWFW and FWF, Austria; ANAS, Azerbaijan; CNPq and FAPESP, Brazil; NSERC, NRC and CFI, Canada; CERN; ANID, Chile; CAS, MOST and NSFC, China; Minciencias, Colombia; MEYS CR, Czech Republic; DNRF and DNSRC, Denmark; IN2P3-CNRS and CEA-DRF/IRFU, France; SRNSFG, Georgia; BMFTR, HGF and MPG, Germany; GSRI, Greece; RGC and Hong Kong SAR, China; ICHEP and Academy of Sciences and Humanities, Israel; INFN, Italy; MEXT and JSPS, Japan; CNRST, Morocco; NWO, Netherlands; RCN, Norway; MNiSW, Poland; FCT, Portugal; MNE/IFA, Romania; MSTDI, Serbia; MSSR, Slovakia; ARIS and MVZI, Slovenia; DSI/NRF, South Africa; MICIU/AEI, Spain; SRC and Wallenberg Foundation, Sweden; SERI, SNSF and Cantons of Bern and Geneva, Switzerland; NSTC, Taipei; TENMAK, T\"urkiye; STFC/UKRI, United Kingdom; DOE and NSF, United States of America.

Individual groups and members have received support from BCKDF, CANARIE, CRC and DRAC, Canada; CERN-CZ, FORTE and PRIMUS, Czech Republic; COST, ERC, ERDF, Horizon 2020, ICSC-NextGenerationEU and Marie Sk{\l}odowska-Curie Actions, European Union; Investissements d'Avenir Labex, Investissements d'Avenir Idex and ANR, France; DFG and AvH Foundation, Germany; Herakleitos, Thales and Aristeia programmes co-financed by EU-ESF and the Greek NSRF, Greece; BSF-NSF and MINERVA, Israel; NCN and NAWA, Poland; La Caixa Banking Foundation, CERCA Programme Generalitat de Catalunya and PROMETEO and GenT Programmes Generalitat Valenciana, Spain; G\"{o}ran Gustafssons Stiftelse, Sweden; The Royal Society and Leverhulme Trust, United Kingdom.

In addition, individual members wish to acknowledge support from CERN: European Organization for Nuclear Research (CERN DOCT); Chile: Agencia Nacional de Investigaci\'on y Desarrollo (FONDECYT 1230812, FONDECYT 1240864, Fondecyt 3240661, Fondecyt Regular 1240721); China: Chinese Ministry of Science and Technology (MOST-2023YFA1605700, MOST-2023YFA1609300), National Natural Science Foundation of China (NSFC - 12175119, NSFC 12275265); Czech Republic: Czech Science Foundation (GACR - 24-11373S), Ministry of Education Youth and Sports (ERC-CZ-LL2327, FORTE CZ.02.01.01/00/22\_008/0004632), PRIMUS Research Programme (PRIMUS/21/SCI/017); EU: H2020 European Research Council (ERC - 101002463); European Union: European Research Council (BARD No. 101116429, ERC - 948254, ERC 101089007), European Regional Development Fund (SMASH COFUND 101081355, SLO ERDF), European Union, Future Artificial Intelligence Research (FAIR-NextGenerationEU PE00000013), Italian Center for High Performance Computing, Big Data and Quantum Computing (ICSC, NextGenerationEU); France: Agence Nationale de la Recherche (ANR-21-CE31-0022, ANR-22-EDIR-0002, ANR-24-CE31-0504-01); Germany: Deutsche Forschungsgemeinschaft (DFG - 469666862, DFG - CR 312/5-2); China: Research Grants Council (GRF); Italy: Istituto Nazionale di Fisica Nucleare (ICSC, NextGenerationEU), Ministero dell'Università e della Ricerca (NextGenEU 153D23001490006 M4C2.1.1, NextGenEU I53D23000820006 M4C2.1.1, NextGenEU I53D23001490006 M4C2.1.1, SOE2024\_0000023); Japan: Japan Society for the Promotion of Science (JSPS KAKENHI JP22H01227, JSPS KAKENHI JP22H04944, JSPS KAKENHI JP22KK0227, JSPS KAKENHI JP24K23939, JSPS KAKENHI JP24KK0251, JSPS KAKENHI JP25H00650, JSPS KAKENHI JP25H01291, JSPS KAKENHI JP25K01023); Norway: Research Council of Norway (RCN-314472); Poland: Ministry of Science and Higher Education (IDUB AGH, POB8, D4 no 9722), Polish National Science Centre (NCN 2021/42/E/ST2/00350, NCN OPUS 2023/51/B/ST2/02507, NCN UMO-2019/34/E/ST2/00393, UMO-2022/47/O/ST2/00148, UMO-2023/49/B/ST2/04085, UMO-2023/51/B/ST2/00920, UMO-2024/53/N/ST2/00869); Portugal: Foundation for Science and Technology (FCT); Spain: Generalitat Valenciana (ASFAE/2022/008), Ministry of Science and Innovation (MCIN \& NextGenEU PCI2022-135018-2, MICIN \& FEDER PID2021-125273NB, RYC2019-028510-I, RYC2020-030254-I, RYC2021-031273-I, RYC2022-038164-I), Ministerio de Ciencia, Innovación y Universidades/Agencia Estatal de Investigaci\'on (PID2022-142604OB-C22); Sweden: Carl Trygger Foundation (Carl Trygger Foundation CTS 22:2312), Swedish Research Council (Swedish Research Council 2023-04654, VR 2021-03651, VR 2022-03845, VR 2022-04683, VR 2023-03403, VR 2024-05451), Knut and Alice Wallenberg Foundation (KAW 2018.0458, KAW 2022.0358, KAW 2023.0366); Switzerland: Swiss National Science Foundation (SNSF - PCEFP2\_194658); United Kingdom: Royal Society (NIF-R1-231091); United States of America: U.S. Department of Energy (ECA DE-AC02-76SF00515), Neubauer Family Foundation.

%
%


%
%
%
%
%
%
%
\section{Data availability}
\atlasdatapolicy{hepdatatext=true, hepdataexist=true, athenatext=true}
\printatlasdatapolicy

\printbibliography
\clearpage
 
\begin{flushleft}
\hypersetup{urlcolor=black}
{\Large The ATLAS Collaboration}

\bigskip

\AtlasOrcid[0000-0002-6665-4934]{G.~Aad}$^\textrm{\scriptsize 103}$,
\AtlasOrcid[0000-0001-7616-1554]{E.~Aakvaag}$^\textrm{\scriptsize 17}$,
\AtlasOrcid[0000-0002-5888-2734]{B.~Abbott}$^\textrm{\scriptsize 122}$,
\AtlasOrcid[0000-0002-0287-5869]{S.~Abdelhameed}$^\textrm{\scriptsize 118a}$,
\AtlasOrcid[0000-0002-1002-1652]{K.~Abeling}$^\textrm{\scriptsize 55}$,
\AtlasOrcid[0000-0001-5763-2760]{N.J.~Abicht}$^\textrm{\scriptsize 49}$,
\AtlasOrcid[0000-0002-8496-9294]{S.H.~Abidi}$^\textrm{\scriptsize 30}$,
\AtlasOrcid[0009-0003-6578-220X]{M.~Aboelela}$^\textrm{\scriptsize 45}$,
\AtlasOrcid[0000-0002-9987-2292]{A.~Aboulhorma}$^\textrm{\scriptsize 36e}$,
\AtlasOrcid[0000-0001-5329-6640]{H.~Abramowicz}$^\textrm{\scriptsize 156}$
% ...
\bigskip
\\

% ...
$^{17}$Department for Physics and Technology, University of Bergen, Bergen; Norway.\\
% ...
$^{30}$Physics Department, Brookhaven National Laboratory, Upton NY; United States of America.\\
% ...
$^{36}$$^{(a)}$Facult\'e des Sciences Ain Chock, Universit\'e Hassan II de Casablanca;$^{(b)}$Facult\'{e} des Sciences, Universit\'{e} Ibn-Tofail, K\'{e}nitra;$^{(c)}$Facult\'e des Sciences Semlalia, Universit\'e Cadi Ayyad, LPHEA-Marrakech;$^{(d)}$LPMR, Facult\'e des Sciences, Universit\'e Mohamed Premier, Oujda;$^{(e)}$Facult\'e des sciences, Universit\'e Mohammed V, Rabat;$^{(f)}$Institute of Applied Physics, Mohammed VI Polytechnic University, Ben Guerir; Morocco.\\
% ...
$^{45}$Physics Department, Southern Methodist University, Dallas TX; United States of America.\\
% ...
$^{49}$Fakult\"{a}t Physik, Technische Universit{\"a}t Dortmund, Dortmund; Germany.\\
% ...
$^{55}$II. Physikalisches Institut, Georg-August-Universit\"{a}t G\"ottingen, G\"ottingen; Germany.\\
% ...
$^{103}$CPPM, Aix-Marseille Universit\'e, CNRS/IN2P3, Marseille; France.\\
% ...
$^{118}$$^{(a)}$New York University Abu Dhabi, Abu Dhabi;$^{(b)}$United Arab Emirates University, Al Ain; United Arab Emirates.\\
% ...
$^{122}$Homer L. Dodge Department of Physics and Astronomy, University of Oklahoma, Norman OK; United States of America.\\
% ...
$^{156}$Raymond and Beverly Sackler School of Physics and Astronomy, Tel Aviv University, Tel Aviv; Israel.\\
% ...
$^{a}$ Also at Affiliated with an institute formerly covered by a cooperation agreement with CERN.\\
$^{b}$ Also at An-Najah National University, Nablus; Palestine.\\
$^{c}$ Also at Borough of Manhattan Community College, City University of New York, New York NY; United States of America.\\
$^{d}$ Also at Center for Interdisciplinary Research and Innovation (CIRI-AUTH), Thessaloniki; Greece.\\
$^{e}$ Also at Centre of Physics of the Universities of Minho and Porto (CF-UM-UP); Portugal.\\
$^{f}$ Also at CERN, Geneva; Switzerland.\\
% ...

\end{flushleft}


%

%
%

%

%
%
%
%

\end{document}